\documentclass[12pt,final]{article}
\usepackage{amsmath,amsfonts,amssymb,color,bm}
\usepackage{mathrsfs} 
\usepackage{epsfig}
\usepackage{psfrag}
\usepackage{upgreek} 
\usepackage{relsize} 
\usepackage{cite}
\usepackage{upgreek}

\newtheorem{theorem}{Theorem}[section]

\newtheorem{proposition}[theorem]{Proposition}

\newtheorem{remark}[theorem]{Remark}

\numberwithin{equation}{section}

\renewcommand\dot[1]{\mathchoice
                 {{\buildrel{\hspace*{.1em}\text{\LARGE.}}\over{#1}}}
                 {{\buildrel{\hspace*{.1em}\text{\Large.}}\over{#1}}}
                 {{\buildrel{\hspace*{.1em}\text{\large.}}\over{#1}}}
                 {{\buildrel{\hspace*{.1em}\text{\large.}}\over{#1}}}}
\newcommand\pdt[1]{\frac{\pl{#1}}{\pl t}}
\newcommand\DT[1]{\mathchoice
   {{\buildrel{\hspace*{.1em}\text{\LARGE.}}\over{#1}}}
   {{\buildrel{\hspace*{.1em}\text{\Large.}}\over{#1}}}
   {{\buildrel{\hspace*{.1em}\text{\large.}}\over{#1}}}
   {{\buildrel{\hspace*{.1em}\text{\large.}}\over{#1}}}}

\newcommand{\wt}[1]{\mathchoice
     {\text{\small$\widetilde{\text{\normalsize$#1$}}\hspace*{.03em}$}}
                    {\text{\small$\widetilde{\text{\normalsize$#1$}}$}}
                    {\widetilde{#1\hspace*{-.02em}}\hspace*{.03em}}
                    {\widetilde{#1}}}
\newcommand{\lineunder}[2]{\LU{\begin{array}[t]{c}\underbrace{#1}\vspace*{.5em}\end{array}}{\mbox{\footnotesize\rm #2}}}
\newcommand{\LU}[2]{\begin{array}[t]{c}#1\vspace*{-1em}\\_{#2}\end{array}}
\newcommand{\linesunder}[3]{\LSU{\begin{array}[t]{c}\underbrace{#1}\vspace*{.5em}\end{array}}{\mbox{\footnotesize\rm #2}}{\mbox{\footnotesize\rm #3}}}
\newcommand{\LSU}[3]{\begin{array}[t]{c}#1\vspace*{-1em}\\_{#2}\vspace*{-.5em}\\_{#3}\end{array}}
\newcommand{\threelinesunder}[4]{\LSUU{\begin{array}[t]{c}\underbrace{#1}\vspace*{.5em}\end{array}}{\mbox{\footnotesize\rm #2}}{\mbox{\footnotesize\rm #3}}{\mbox{\footnotesize\rm #4}}}
\newcommand{\LSUU}[4]{\begin{array}[t]{c}#1\vspace*{-1em}\\_{#2}\vspace*{-.5em}\\_{#3}\vspace*{-.5em}\\_{#4}\end{array}}

\def\R{{\mathbb R}}

\def\bbA{{\mathbb A}} \def\bbB{{\mathbb B}} 
\def\bbD{{\mathbb D}}  
  \def\bbI{{\mathbb I}}
\def\bbJ{{\mathbb J}} \def\bbK{{\mathbb K}} 
\def\bbM{{\mathbb M}}

\def\FG{\boldsymbol}
\def\aa{{\FG a}} \def\bb{{\FG b}}  
 \def\ee{{\FG e}}  
  
\def\jj{{\FG j}}   
 \def\nn{{\FG n}} \def\oo{{\FG o}} 
\def\pp{{\FG p}} \def\qq{{\FG q}}  
   
\def\vv{{\FG v}} \def\ww{{\FG w}} \def\xx{{\FG x}} 
\def\yy{{\FG y}}  
   
\def\DD{{\FG D}} 
\def\FF{{\FG F}} 
 
\def\GG{{\FG G}}  
  \def\LL{{\FG L}} 
   
 \def\QQ{{\FG Q}}  
\def\SS{{\FG S}} \def\TT{{\FG T}} 
\def\VV{{\FG V}}  \def\XX{{\FG X}} 
 
\def\ALPHA{\mathsf a} \def\BETA{\mathsf b} \def\GAMMA{\mathsf c}

\newcommand\SYM{\R_\mathrm{sym}^{d\times d}}

\renewcommand\d{\mathrm d}

\newcommand{\KAPA}{\mbox{\Large$\upkappa$}}
\newcommand{\kapa}{\mbox{\Large$\kappa$}}
\newcommand{\Kapa}{\mbox{\large$\kappa$}}

\newcommand{\GRAVITY}{{\bm g}}

\begin{document}
\begin{sloppypar}

\pretolerance=10000  

\def\rmn{{\rm n}}
\def\rmt{{\rm t}}
\def\pl{{\partial}}
\def\In{{\in}}
\newcommand{\eq}[1]{\eqref{#1}}
\def\uhalf{{u_\tau^{k-1/2}}}
\def\vhalf{{v_\tau^{k-1/2}}}
\def\qhalf{{q_\tau^{k-1/2}}}
\def\phalf{{p_\tau^{k-1/2}}}
\def\zhalf{{z_\tau^{k-1/2}}}
\def\pihalf{{\pi_\tau^{k-1/2}}}
\newcommand{\QED}{\ \hfill$\ \hfill\Box$\medskip}
\newcommand{\overf}{\hspace*{.08em}\overline{\hspace*{-.08em}f}}
\def\eps{\varepsilon}
\def\bbI{\mathbb I}
\def\bbM{\mathbb M}
\def\dev{\mathrm{dev}}
\def\DEV{\R_\mathrm{dev}^{d\times d}}
\def\SYM{\R_\mathrm{sym}^{d\times d}}
\newcommand{\Fe}{\FF_{\hspace*{-.1em}\mathrm e^{^{^{}}}}}
\newcommand{\Fezero}{\FF_{\hspace*{-.1em}\mathrm e,0}^{}}
\newcommand{\Fp}{\FF_{\hspace*{-.1em}\mathrm p^{^{^{}}}}}
\newcommand{\Fpzero}{\FF_{\hspace*{-.1em}\mathrm p^{^{^{}}},0}}
\newcommand{\DTFe}{\DT\FF_{\!\mathrm e^{^{^{}}}}}
\newcommand{\DTFp}{\DT\FF_{\!\mathrm p^{^{^{}}}}}
\newcommand{\Fee}{\FF_{\mathrm e}}
\newcommand{\Fpp}{\FF_{\mathrm p}}
\newcommand{\Lp}{\LL_{\mathrm p^{^{^{}}}}}
\newcommand{\mfL}{\mathfrak L}
\newcommand{\barOmega}{\,\overline{\!\varOmega}}

\def\ZETA{z}

\newcommand{\ol}[1]{\overline{#1}}
\newcommand{\ds}[1]{\displaystyle{#1}}
\newcommand{\bma}{\left(\begin{array}}
\newcommand{\ema}{\end{array}\right)}
\newcommand{\plt}[1]{\pdt{#1}}

\def\Dv{\bfD}
\def\Lv{\LL}

\vspace*{2cm}

\baselineskip=20pt\noindent
{\large\bf 
A general thermodynamical model for finitely-strained continuum
with inelasticity and diffusion, its {\smaller GENERIC} derivation in Eulerian
formulation, and some application.
}
\bigskip

\noindent
      {\sc Alexander Mielke}$\,^{1}$
and
      {\sc Tom{\'a}{\v s} Roub{\'\i}{\v c}ek}$\,^{2,3}$ 

\bigskip
\bigskip
\baselineskip=10pt

\noindent{\footnotesize\it
$^1$ Weierstrass-Institut f\"ur Angewandte Analysis und 
Stochastik, Mohrenstr.39, D-10117 Berlin, Germany
\\
$^2$ Mathematical Institute, Charles University,
Sokolovsk{\'a}~83, CZ-186~75~Praha~8, Czech Republic.\\
$^3$ Institute of Thermomechanics, Czech Acad.\ Sci., Dolej\v skova~5,
CZ--182~00 Praha 8, Czech Republic}

\bigskip
\bigskip

\noindent\hspace*{1cm}
\begin{minipage}[t]{14.3cm}

\baselineskip=12pt

{\small
\noindent{\bf Abstract.} 
A thermodynamically consistent visco-elastodynamical model at finite
strains is derived that also allows for inelasticity (like plasticity or
creep), thermal coupling, and poroelasticity with diffusion.  The theory
is developed in the  Eulerian framework  and is shown to be consistent
with the thermodynamic framework given by  General Equation for
Non-Equilibrium Reversible-Irreversible Coupling ({\smaller GENERIC}). For
the latter we use that the transport terms are given in terms of Lie
derivatives.  Application is illustrated by two examples,
namely  volumetric phase transitions with dehydration in rocks
and martensitic phase transitions  in shape-memory alloys. A strategy
towards a rigorous mathematical analysis is only very briefly outlined.

\medskip


\medskip

} 
\end{minipage}

\def\GM{G_\text{\sc m}^{}}
\def\GMxi{G_\text{\sc m}^{\bm\xi}}
\def\GMxikk{G_{\text{\sc m}^{}}^{\bm\xi_\tau^{k-1}}}
\def\GMxitau{G_{\text{\sc m}^{}}^{\underline{\bm\xi}_\tau}}
\def\W{w}
\def\COUPLING{\gamma}
\def\OMEGA{\omega}
\def\bfb{\bb}
\def\bfo{\oo}
\def\bfj{\jj}
\def\rmR{\text{\sc r}}
\def\bfF{\FF}
\def\bfL{\LL}
\def\bfT{\bfSigma_\mathrm{Cauchy}}
\def\bfQ{\QQ}
\def\bfD{\DD}
\def\bfV{\VV}
\def\bfSigma{\bm\varSigma}
\def\bfxi{{\bm\xi}}
\def\bfzeta{{\bm\zeta}}
\def\bfXi{{\bm\Xi}}
\def\rmp{{\rm p}}
\def\rmD{{\rm D}}
\def\rmT{{\rm T}}
\def\rmA{{\rm A}}
\def\rmC{{\rm C}}
\def\rmB{{\rm B}}
\def\rmE{{\rm E}}
\def\rmL{{\rm LL}}
\def\DIV{{\rm div}}
\def\bfeta{{\bm\eta}}
\def\calE{{\mathcal E}}
\def\calS{{\mathcal S}}
\def\calR{{\mathcal R}}
\def\rmd{{\d}}
\def\ti{\times}
\def\bfXi{{\bm\varXi}}

\allowdisplaybreaks
 
\vspace*{1em}
\baselineskip=16pt

\section{Introduction}\label{sect-intro}

In this article, we give a systematic derivation of a fairly
general visco-elastodynamical model with possibly certain internal variables at large
strains (also called finite strain) in the Eulerian setting. The
mechanical attributes can be summarized as follows:
\begin{enumerate}\itemsep-0.1em
\item[(i)] hyperelasticity for the elastostatic part (i.e.\ the conservative
  part of the Cauchy stress comes from a free energy $\psi$),

\item[(ii)] Jeffreys' rheology (also called anti-Zenner rheology) with the
  multiplicative de\-compo\-si\-tion of the deformation gradient into the elastic and
  the inelastic (plastic/creep) distortions,

\item[(iii)] Fick-type diffusion of an intensive variable (like phase
  field),

\item[(iv)] Fick-type diffusion of an extensive variable (like a conserved
  chemical species), 

\item[(iv)] and the heat equation (either as an energy- or an entropy-balance
  equation).
\end{enumerate}
Our goal is to devise a model in the Eulerian formulation, consistent with
sound thermodynamical principles of energy conservation, entropy
entropy-production balance, an non-negativity of temperature. These
properties might still be satisfied by various other models, hence we
show that our class of models
also fits into the so-called  {\smaller GENERIC}
framework, which is the acronym standing for {\it General Equation for
  Non-Equilibrium Reversible-Irreversible Coupling}. This name which was
introduced in \cite{GrmOtt97DTCF12} but this class of models
has its origins in the metriplectic theory developed in \cite{Morr84BFIC,Morr86PJHD},
cf.\ the survey \cite{Morr09TBDO}.  Over the last decade, the {\smaller GENERIC}
framework has proved to be a versatile modeling tool for various complex coupled models
for fluids and solids, see e.g.\ \cite{LeJoCa08UNET, Miel11FTDM, HutSve12TMFV,
  DuPeZi13GFVF, PaKlGr18MTDI, PaPeKl20HCM,BetSch19EMEC, Lasa21ATCN, PTPS22CNCF,
  ZaPeTh23GFRF} and the references therein.  We also refer to
\cite{MiPeZi24DGSH} for a derivation of a dissipative {\smaller GENERIC} system
from (non-dissipative) Hamiltonian systems.\smallskip

The plan of this paper is as follows: In Sect.~\ref{sec-kinem} we recall 
the standard Eulerian kinematics in continuum mechanics with Eulerian
velocity $\vv$. Sect.\,\ref{sec-system} is devoted to the formulation of 
a rather general model for compressible
thermo-visco-elastodynamics with inelasticity motivated by the multiplicative
split $\FF = \Fe \Fp$ but replaced by the kinematic equation 
\[
\dot \Fe =(\nabla \vv)\Fe - \Fe\,\Lp\,,
\]
where the elastic part $\Fe\in \R^{d\ti d}$ of the deformation will be a state
variable while the inelastic distortion rate tensor $\Lp$ will be given in
terms of a flow rule involving the Mandel tensor. Additionally, we allow for
the diffusion of an intensive variable or extensive variable $z$. The model is
phrased in the classical mechanical approach using the heat equation in terms
of the absolute temperature
$\theta>0$ and the referential free energy $\uppsi(\Fe,z,\theta)$. 

To give a physically more sound justification, 
in Sect.\,\ref{sec-GENERIC} we do a step aside and derive a quite similar
general continuum model, but start from a completely different angle. Instead
of studying balance equations and constitutive laws, we follow the philosophy
of {\smaller GENERIC} where the model is determined by an energy and an entropy
functionals $\calE$ and $\calS$  as well as geometric operators $\bbJ$ and
$\pl\calR^*$ describing the Hamiltonian and the dissipative parts of the
evolution, namely
\[
\frac{\pl}{\pl t} q = \bbJ(q) \rmD \calE(q) +
\pl_\xi\calR^*\big(q,\rmD\calS(q)\big). 
\]
A particular advantage of this theory is that it easily allows for coordinate
changes and consistent coupling of different effects, see \cite{Miel11FTDM,
  HutSve12TMFV, PaKlGr18MTDI, PaPeKl20HCM, PTPS22CNCF, ZaPeTh23GFRF}. In
particular, it is useful that one is able to choose an arbitrary
thermodynamical variable $w$ (e.g.\ the internal energy $e$, the entropy $s$,
the temperature $\theta$, or its inverse $1/\theta$) when deriving the energy
balance or the entropy imbalance. Using $e=E(\Fe,z,w)$ and $s=S(\Fe,z,w)$ one
has
\[
\theta =\Theta(\Fe,z,w)= \frac{E'_w(\Fe,z,w)}{S'_w(\Fe,z,w)} \quad \text{and}
\quad \Sigma_\text{Cauchy} =\big[\, (E'_{\Fe}{-} \Theta S'_{\Fe}) \Fe^\top + (E{-}\Theta
S)\bbI \,\big]_{(\Fe,z,w)}
\]
for all choices of $w$. As a natural by-product, the {\smaller GENERIC}
structure reveals in a transparent way which source terms appear in the energy
equation and which ones in the entropy equation.  To the best of our knowledge,
this provides the first complete treatment of 
Eulerian elasticity in {\smaller GENERIC}. 

In  Sect.\,\ref{sec-deriv}, we study the impact of the {\smaller GENERIC} 
formulation on the model developed in Sect.\,\ref{sec-system} formulated
in terms of temperature and referential free energy. In
Sect.~\ref{sec-appl}, we illustrate the possible application on two examples
involving volumetric and spherical phase transitions, namely
Earth’s mantle dynamics with (de)hydration and martensitic phase transition
with plasticity and possibly also a metal-hydrid phase transition. 
Finally, we comment (mostly very technical) analytical aspects very briefly and
only conceptually in Sect.~\ref{sect-notes}.

\medskip

For readers' convenience, let us summarize 
the basic notation used in what follows:
\begin{center}
\fbox{
\begin{minipage}[t]{16em}\small\smallskip
$\yy$ deformation,\\
  $\vv$ velocity,\\
  $\Lv=\nabla\vv$ velocity gradient,\\
$\varrho$ mass density,\\
$\bfSigma_\mathrm{Cauchy}$ the Cauchy stress,\\
$\bfSigma_\mathrm{Mandel}$ the Mandel stress,\\
$\bfSigma_\mathrm{dissip}$ the dissipative stress,\\
$\FF$ deformation gradient,\\
$\Fe$ elastic strain,\\
$\Fp$ inelastic strain,\\
$\Lp$ inelastic distortion rate,\\
 $\ZETA$ content of diffusant ($\alpha$ or $\beta$),\\
  $\mu$ chemical potential,\\
  $\theta$ temperature,\\
$e$ internal energy ($=\psi+\theta s$),\\
$\psi$ free energy (actual),\\
$\uppsi$ free energy (referential),\\
$s$ entropy,\\
$\bm\xi$ return mapping,
\\
$\Dv={\rm sym}\Lv=(\Lv^\top\!{+}\Lv)/2$,\\
$\rho_\text{\sc r}$ referential mass density,\\
$\GRAVITY$ gravity acceleration,
\end{minipage}
\begin{minipage}[t]{19em}\small\smallskip
$J=\det\FF$ Jacobian\,=\,determinant of $\FF$,\\
$p_{\rm heat}$ heat production rate,\\
$p_{\rm mech}$ mechanical power,\\
$\sigma_{\rm prod}$ entropy production rate,\\
$\jj_{\rm ener}$ energy flux,\\
$\jj_{\rm entr}$ entropy flux,\\
$\bbK_\mathrm{heat}$ heat conductivity,\\
$\bbK_\mathrm{diff}$ diffusivity/mobility
($\bbA_\mathrm{diff}$ or $\bbB_\mathrm{diff}$),\\
$R_{\rm plast}$ inelastic entropy-production potential,\\
$A_\mathrm{source}$ source of intensive variable $\alpha$,\\
  tr$(\cdot)$ trace of a matrix,\\
  dev$(\cdot)$ deviatoric part of a matrix\\
$\R_{\rm sym}^{d\times d}$ set of symmetric matrices,\\
$\R_{\rm dev}^{d\times d}=\{A\in\R_{\rm sym}^{d\times d};
\ {\rm tr}A=0\}$,\\
$\bbJ$ Poisson operator in {\smaller GENERIC},\\
$\bbK$ Onsager operator in {\smaller GENERIC},\\
$w$ general thermal variable (e.g.\ $e,\theta,1/\theta,s$),\\
$E(\Fe,\alpha,\beta,w)$ internal energy as function,\\
$S(\Fe,\alpha,\beta,w)$ entropy as function,\\ 
$\mfL_\vv\Box$ Lie derivative w.r.t.\ the vector field $\vv$,
\\
$\Box$ a general placeholder (or the end of proofs).
\smallskip \end{minipage}
}\end{center}

\vspace{-.9em}

\begin{center}
{\small\sl Table\,1.\ }
{\small
Summary of the basic notation used. 
}
\end{center}

\section{Kinematics}\label{sec-kinem}

In the finite-strain (also called large-strain) continuum mechanics, the basic
geometrical concept is a {\it deformation} $\yy:\Omega\to\R^d$ as a mapping from a
reference configuration $\Omega\subset\R^d$ into the physical space $\R^d$.
The inverse motion $\bm\xi=\yy^{-1}:\yy(\Omega)\to\Omega$, if it
exists, is called a {\it return } (or sometimes a {\it reference})
{\it mapping}. We will denote by $\XX$ and $\xx$ the reference (Lagrangian) and
the actual (Eulerian) point coordinates, respectively. The other basic geometrical
object is the (referential) {\it deformation gradient}
$\FF_\text{\!\sc r}^{}(\XX)=\nabla_{\XX}^{}\yy$.

If evolving in time, $\xx=\yy(t,\XX)$ is sometimes called a ``motion''.
The important quantity is the (referential)
velocity $\vv_\text{\sc r}^{}=\frac{\d}{\d t}\yy(t,\XX)$
with $\d/\d t$ the derivative with respect to time of a time dependent
function. When composed with the return mapping $\bm\xi$, we obtain
the Eulerian representations 
\begin{align}\label{F-v-Eulerian}
\FF(t,\xx)=\FF_\text{\!\sc r}^{}(t,\bm\xi(\xx))\ \ \ \ \text{ and }\ \ \ \
\vv(t,\xx)=\vv_\text{\!\sc r}^{}(t,\bm\xi(\xx))\,.
\end{align}
The Eulerian velocity $\vv$ is employed in the convective time derivative
\begin{align}
(\bm\cdot)\!\DT{^{}}=\frac{\pl}{\pl t}(\bm\cdot)+(\vv{\cdot}\nabla)(\bm\cdot)
\end{align}
with $\nabla$ taken with respect to actual
coordinates, to be used for scalars and, component-wise, for vectors or tensors.

Then the velocity gradient
$\nabla\vv=\nabla_{\!\XX}^{}\vv\nabla_{\!\xx}^{}\XX=\DT\FF\FF^{-1}$,
where we used the chain-rule calculus  and
$\FF^{-1}=(\nabla_{\!\XX}^{}\xx)^{-1}=\nabla_{\!\xx}^{}\XX$. This gives the
{\it transport-and-evolution equation} the so-called {\it kinematic equation})
for the deformation gradient as
\begin{align}
\DT\FF=\Lv\FF\ \ \ \text{ with }\ \Lv:=\nabla\vv\,.
\label{ultimate}\end{align}
From this, we also obtain the kinematic equation for 
$\det\FF$ as $\DT{\overline{\det\FF}}=(\DIV\,\vv)\det\FF$.

Introducing a (generally non-symmetric) {\it inelastic distortion} tensor $\Fp$,
a conventional large-strain plasticity is based on Kr\"oner-Lie-Liu
\cite{Kron60AKVE,LeeLiu67FSEP} {\it multiplicative
decomposition}\index{multiplicative decomposition}
\begin{align}\label{Green-Naghdi}
\FF=\Fe\Fp\,.
\end{align}
Here $\Fe=\FF\Fp^{-1}$ is the {\it elastic distortion}. The interpretation of
$\Fp$ is a transformation of the reference configuration into an intermediate
stress-free configuration, and then $\Fe$ transforms this intermediate
configuration into the current actual configuration.

Applying the material derivative on \eq{Green-Naghdi} and using
\eq{ultimate}, we obtain $\Lv\FF=\DT\FF=\DTFe\Fp+\Fe\DTFp$ and,
multiplying it by $\FF^{-1}=\Fpp^{-1}\Fee^{-1}$, we eventually obtain
\begin{align}
\Lv=\!\!\!\!\!\!\threelinesunder{\DTFe\Fee^{-1}\!\!\!}{elastic\ \ }{distortion}{rate}\!\!\!+\Fe\!\!\!\!\!\!\!\threelinesunder{\DTFp\Fpp^{-1}\!\!\!\!}{inelastic}{distortion}{rate $=:\Lp$}\!\!\!\!\Fee^{-1},
\label{dafa-formula}\end{align}
which is used mostly in connection to plasticity, cf.\ e.g.\
\cite{BesGie94MMID,Dafa84PSCS,GurAna05DMSP,GuFrAn10MTC,JirBaz02IAS,Lee69EPDF,RajSri04TMMN};
the term ``distortion rates'' is due to \cite{GurAna05DMSP,GuFrAn10MTC}
while sometimes $\Lp$ is called a ``plastic dissipation tensor''
\cite{BesGie94MMID} or  ``velocity gradient of purely plastic deformation''
in \cite{Lee69EPDF}, etc. By the algebraic manipulation, we can eliminate
$\Fp$ and see that \eq{dafa-formula} is equivalent to the
{\it kinematic equation} for $\Fe$:
\begin{align}
\DT\Fe=\Lv\Fe-\Fe\Lp\,.
\label{evol-of-E}\end{align}
In principle, if one is interested also in the inelastic distrotion itself,
we can reconstruct $\Fp$ from \eq{dafa-formula} when re-arranging it to the
plastic-distortion evolution rule $\DT\Fp=\Lp\Fp$ and by prescribing an initial
condition $\Fp|_{t=0}^{}$.

\section{A thermo-visco-elastodynamics with diffusion}\label{sec-system}

To come straight to a quite general model, which represents a
concrete motivation for the next section. We first formulate
it in a manner which is quite common in engineering and
physics, specifically using a {\it free energy} from which one can
read both the internal energy and the entropy. Moreover, it
is quite standard to use the {\it referential} free energy
$\uppsi=\uppsi(\Fe,\ZETA,\theta)$ considered in J/m$^{-3}$=Pa,
i.e.\ a specific energy per referential volume,
and the heat equation formulated in terms of temperature.
For another, also a standard setting involves a referential free energy
considered in J/kg, see Remark~\ref{rem-Einstein} below.

The other ingredient is the dissipation potential
$r=r(\ZETA,\theta;\cdot,\cdot):\R^{d\times d}\times\R_{\rm dev}^{d\times d}\to\R$
acting on the velocity gradient $\LL$ (or here rather only
on its symmetric part) and the inelastic distortion rate $\Lp$.

\def\SW{\mathfrak{s}_\text{\sc i}^\text{\sc e}}
\def\SW{\mathfrak{s}_\text{\sc ext}^{}}

For the sake of generality, we distinguish two cases concerning the
content or concentration variable $\ZETA$: {\it extensive} or {\it intensive},
later in Section~\ref{sec-GENERIC} denoted respectively by $\alpha$ and $\beta$,  
the former case leading to some additional terms. To write the equations
more ``compactly'' for both cases, we will use the ``switch''
\begin{align}\label{switch}
\SW=
\begin{cases}1&\text{if $\ZETA$ is intensive},\\[-.2em]
0&\text{if $\ZETA$ is extensive}.
\end{cases}
\end{align}
This affects the conservative Cauchy stress $\bfSigma_\mathrm{Cauchy}$ in
\eq{momentum-eq} and the diffusion equation \eq{system-diffusion-eq} as well as
the temperature equation \eq{Euler-large-heat-eq} below.  The system for the
six-tuple $(\varrho,\vv,\Fe,\ZETA,\Lp,\W)$ is composed from the six equations,
namely the mass-density continuity equation, the momentum equation, the
kinematic equation \eq{evol-of-E}, the diffusion equation, the inelastic flow
rule, and the heat equation:
\begin{subequations}\label{the-system}
\begin{align}\label{cont-eq}
&\pdt\varrho+\DIV(\varrho\vv)=0
\\&\nonumber  \pdt{}(\varrho \vv)+\DIV(\varrho\vv{\otimes}\vv)
=\varrho\GRAVITY+\DIV\big(\bfSigma_\mathrm{Cauchy}
{+}\bfSigma_\mathrm{dissip}\big)\ \ \text{ with }\
\bfSigma_\mathrm{dissip}=[r^*]_\Dv'(\ZETA,\theta;\Dv,\Lp)
\\&\hspace{11em}\text{and }\ 
\bfSigma_\mathrm{Cauchy}
=\dfrac{\uppsi_{\Fe}'(\Fe,\ZETA,\theta)\Fe^\top\!{-}\,\SW\ZETA\uppsi_\ZETA'(\Fe,\ZETA,\theta)\bbI}{\det\Fe}\,,
\label{momentum-eq}
\\&\DT\Fe=(\nabla\vv)\Fe-\Fe\Lp\,,
\label{evol-of-E+}
\\&
\DT\ZETA+\SW\ZETA\,\DIV\,\vv=
\DIV\Big(\bbK_\mathrm{diff}\nabla\dfrac\mu\theta\Big)
\label{system-diffusion-eq}
\ \ \ \text{ with }\
\mu=\frac{\uppsi_\ZETA'(\Fe,\ZETA,\theta)}{\det\Fe}\,,
\\&r_{\Lp}'\!(\ZETA,\theta;\Dv,\Lp)={\rm dev}\,\bfSigma_\mathrm{Mandel}
\ \ \ \text{ with }\
\bfSigma_\mathrm{Mandel}=\frac{\Fe^\top\uppsi_{\Fe}'(\Fe,\ZETA,\theta)}{\det\Fe}\,,
\label{inelastic-flow}
\\&\nonumber\hspace*{-0em}
c(\Fe,\ZETA,\theta)\DT\theta
=p_\mathrm{heat}-\DIV\Big(\bbK_\mathrm{heat}\nabla\frac1\theta\Big)+\theta
\frac{\uppsi_{\Fe\theta}''(\Fe,\ZETA,\theta)\Fe^\top\!\!}{\det\Fe}{:}\nabla\vv
-\theta\frac{\Fe^\top\uppsi_{\Fe\theta}''(\Fe,\ZETA,\theta)}{\det\Fe}{:}\Lp
\\&\nonumber\hspace{13em}
-\frac{\uppsi_\ZETA'(\Fe,\ZETA,\theta){-}\theta\uppsi_{\ZETA\theta}''(\Fe,\ZETA,\theta)\!}{\det\Fe}\,\DT\ZETA-\SW\frac{\!\ZETA\uppsi_\ZETA'(\Fe,\ZETA,\theta)\!}{\det\Fe}\,\DIV\,\vv
\\&\nonumber\hspace*{5em}
\text{ with the heat production rate }\ p_\mathrm{heat}=\bfSigma_\mathrm{dissip}{:}\Dv+r_{\Lp}'\!(\ZETA,\theta;\Dv,\Lp){:}\Lp
\\[-.2em]&\hspace{5em}
\text{ and the heat capacity }\ \ c(\Fe,\ZETA,\theta)=-\theta\frac{\uppsi_{\theta\theta}''(\Fe,\ZETA,\theta)}{\det\Fe}\,,
\label{Euler-large-heat-eq}
\end{align}
\end{subequations}
where $\bbK_\mathrm{diff}=\bbK_\mathrm{diff}(\Fe,\ZETA,\theta)$ and
$\bbK_\mathrm{heat}=\bbK_\mathrm{heat}(\Fe,\ZETA,\theta)$
are the (symmetric positive definite) matrices of diffusivity (mobility) and
heat conductivity coefficients, and where $r^*(\ZETA,\theta;\cdot,\cdot)$
denotes the convex conjugate to the dissipation potential
$r(\ZETA,\theta;\cdot,\cdot)$. This system will be derived and
thermodynamically justified in Section~\ref{sec-deriv} by exploiting a
universal tool in the next Section~\ref{sec-GENERIC}.

To ensure non-negativity of temperature (sometimes called the 0$^{\rm th}$ law of
thermo\-dynamics), the $\uppsi_\ZETA'(\Fe,\ZETA,\theta)\DT\ZETA$-term suggests the
restriction $\uppsi_\ZETA'(\Fe,\ZETA,0)=0$. A particular and perhaps physically
most relevant ansatz to satisfy $\uppsi_\ZETA'(\Fe,\ZETA,0)=0$ is
\begin{align}\label{one-ansatz}
\uppsi(\Fe,\ZETA,\theta)=\theta\upeta(\Fe,\ZETA)+\upphi(\Fe,\theta)\,.
\end{align}
With this ansatz, the term
$(\uppsi_\ZETA'(\Fe,\ZETA,\theta){-}\theta 
 \uppsi_{\ZETA\theta}''(\Fe,\ZETA,\theta))\,\DT\ZETA$
vanishes identically and also the heat capacity
$c(\Fe,\ZETA,\theta)=-\theta\uppsi_{\theta\theta}''(\Fe,\ZETA,\theta)/\!\det\Fe
=-\theta\upphi_{\theta\theta}''(\Fe,\theta)/\!\det\Fe$
becomes independent of $\ZETA$, so that the diffusant content
does not affect the heat capacity and, altogether,
the diffusion does not directly affect the heat equation
at all. Anyhow, there is some experience
that, in some situations, diffusion can directly generate heat so that
the ansatz \eq{one-ansatz} may be not entirely universal.

\section{A general setup of the model for {\smaller GENERIC}} 
\label{sec-GENERIC}

We consider a viscoelastoplastic material in an Eulerian domain $\varOmega$,
which is further characterized by the local temperature $\theta$, an intensive
variable $\alpha$ (like damage or aging or concentration of a diffusant)
and an extensive variable $\beta$ (like a content of a diffusant).
The material properties are encoded in the {\it actual free-energy density} 
\begin{align}\label{psi}
\psi=\psi(\Fe,\alpha,\beta,\theta)
\end{align}
which is related with the referential free energy $\uppsi$ used in
Section~\ref{sec-system} by
\begin{align}\label{psi-vs-psi}
\psi(\Fe,\ZETA,\theta)=\frac{\uppsi(\Fe,\ZETA,\theta)}{\det\Fe}\,.
\end{align}
Both $\psi$ and $\uppsi$ are in the physical unit Pa, i.e.\ J/m$^3$,
but m$^3$ is meant as an actual volume vs a referential volume, respectively.
The latter option can alternatively be considered in J/kg, cf.\
Remark~\ref{rem-Einstein} below.
For consistency with {\smaller GENERIC} to be studied below, we  introduce
the entropy density $S$ and the internal-energy density $E$ via 
\[
S(\Fe,\alpha,\beta,\theta)=- \psi'_\theta(\Fe,\alpha,\beta,\theta) \quad 
\text{and} \quad   E(\Fe,\alpha,\beta,\theta) = \psi(\Fe,\alpha,\beta,\theta) +
\theta S(\Fe,\alpha,\beta,\theta).
\]

The system we want to study will be formulated in terms of the state
$q=(\pp,\Fe,\alpha,\beta,\theta)$, where $\pp=\varrho\vv$ is the linear
momentum and 
\begin{align}\label{rho}
\varrho=\frac{\rho_\text{\sc r}}{\det\Fe}\,.
\end{align}
Here $\Fe$ is the elastic part of the deformation-gradient tensor
$\bfF=\Fe\bfF_\rmp$, viz \eq{ultimate}, but $\bfF_\rmp$ is not needed while
only the inelatic distortion rate $\Lp$ will appear, viz
\eq{dafa-formula}--\eq{evol-of-E}. 

Assuming $\mathrm{tr}\Lp=0$, the system for $(\vv,\Fe,\alpha,\beta,\theta)$
takes the following form:
\begin{subequations}\label{system++}\begin{align}
  \plt{}(\varrho\vv) + \DIV(\varrho\vv{\otimes}\vv)
  &=\DIV\big(\bfSigma_\mathrm{Cauchy}-\beta \psi'_\beta\bbI +
  \bbD_\mathrm{visc}(q) \Dv \big) \,,
  \\
  \plt{\Fe}+(\vv{\cdot}\nabla)\Fe &=(\nabla\vv)\Fe- \Fe \Lp\,,
  \\\label{system++b}
  \plt\alpha +\vv{\cdot}\nabla \alpha &=\DIV\Big( \bbA_\mathrm{diff} \nabla 
  \frac{\mu^\alpha}\theta\Big) - A_\mathrm{source}\frac{\mu^\alpha}\theta\,
  \\
  \plt \beta+\DIV(\beta\vv)&= 
  \DIV\Big(\bbB_\mathrm{diff} \nabla\frac{\mu^\beta}\theta\Big) \,\\
\plt e+ \DIV(e\vv) &=\big(\bfSigma_\mathrm{Cauchy}- \beta\psi'_\beta \bbI +
\bbD_\mathrm{visc} \Dv\big){:}\Dv - \DIV \Big(\bbK_\mathrm{heat} \nabla
\frac1\theta\Big)\,, 
\intertext{where we have}
&\hspace{-5em}\bfSigma_\mathrm{Cauchy}=\psi'_{\Fe} \Fe^\top + \psi  \bbI,\quad
\Dv=\frac12\big(\nabla\vv {+}(\nabla\vv)^\top\big), \quad \mu^\alpha =
\psi'_\alpha, \quad \mu^\beta=\psi'_\beta,
\\&\hspace{-5em}
\Lp= \theta \bbM^*\pl_{\bfeta^{\Fe}} R^*_\mathrm{plast} \big(q, \bbM(q)
\bfSigma_\mathrm{Mandel}\big) \quad \text{ with }\bfSigma_\mathrm{Mandel} =
\Fe^\top \psi'_{\Fe}\,,\ \ \text{ and}
\label{system++Mandel}\\&\hspace{-5em}
e=E(\Fe,\alpha,\beta,\theta)\,,\ \ \text{ and }\ \varrho=\rho_\rmR/\!\det\Fe
\ \ \text{ as in \eq{rho}}\,.
\end{align}\end{subequations}
On the left-hand sides of (\ref{system++}a-e), we can see
the appropriate convective derivatives, which are in fact
Lie derivatives, see Proposition \ref{pr:LieDeriv}. The coefficient
$A_\mathrm{source}=A_\mathrm{source}(\Fe,\ZETA,\theta)$ in the
intensive-variable evolution \eq{system++b} can be used for modeling
damage- or aging-type processes.

The aim of the following section is to show that this system can be cast in the
{\smaller GENERIC} framework.

\subsection{The principles of {\smaller GENERIC}}\label{su:GEN.princ}

We consider states $q$ in a state space $\bfQ$ which is either a flat space or a
smooth manifold. A {\smaller GENERIC} system is a quintuple $(\bfQ,\calE,\calS,\bbJ,\bbK)$,
where the energy $\calE$ and the entropy $\calS$ are differentiable functions
on $\bfQ$ with differentials $\rmD\calE(q),\rmD\calS(q)\in \rmT_q\bfQ$. The
geometric structures are the Poisson operator $\bbJ$ for Hamiltonian dynamics
and the Onsager operator $\bbK$  for dissipative dynamics, which maps $\rmT^*\bfQ $ to
$\rmT \bfQ$. The evolution equation then takes the form 
\[
\plt q = \bbJ(q) \rmD\calE(q) + \bbK(q) \rmD \calS(q). 
\]
The Poisson operator is defined by being skew-symmetric and safisfying the
Jacobi identity, i.e.\ 
\begin{equation}
  \label{eq:GEN.Jacobi}
  \big\langle \zeta_1, \rmD\bbJ(q)[\bbJ(q)\zeta_2] \zeta_3\big\rangle +
\text{cycl.\,perm.}\equiv 0\ \ \text{ for all }\ \zeta_1,\zeta_2,\zeta_3 \in \rmT^*_q\bfQ.
\end{equation}
The Onsager operators are defined by the conditions of symmetry and positive
semi-definiteness, namely $\bbK(q)^*=\bbK(q)\geq 0$. 

The main condition for {\smaller GENERIC} are the so-called \emph{non-interaction
  conditions}, namely 
\begin{equation}
\label{eq:GENERIC.NIC}
\bbJ(q) \rmD\calS(q)\equiv 0 \qquad 
\text{ and }\qquad\bbK(q)\rmD\calE(q)\equiv 0\,. 
\end{equation}
Using the chain rule, a simple consequence of the second condition is energy
conservation along solutions, i.e.\ $\frac\rmd{\rmd t} \calE(q(t))=0$, while
the second condition implies entropy increase, namely $\frac\rmd{\rmd t}
\calS(q(t))=\langle \rmD\calS(q),\bbK(q)\rmD\calS(q)\rangle \geq 0$. See
\cite[Sec.\,2.2]{Miel11FTDM} and \cite{Otti05BET} for further properties of
{\smaller GENERIC} systems. 

In fact, often the linear kinetic relation $\zeta \mapsto \bbK(q)\zeta$ for the
dissipative part needs to be generalized to allow for nonlinear relations. In
such cases, one uses the dual dissipation potential $\calR^*:\rmT^*\bfQ \to
[0,\infty]$, where $\calR^*(q,\cdot):\rmT^*_q \to [0,\infty]$ is a lower
semicontinuous and convex functional satisfying $0=\calR^*(q,0)\leq
\calR^*(q,\zeta)$. In the linear form it takes the form
$\calR^*(q,\zeta)=\frac12\langle \zeta,\bbK(q)\zeta\rangle$.  
Then, the kinetic relation takes the form $\zeta \mapsto
\pl_\zeta\calR^*(q,\zeta) \subset \rmT \bfQ$, where $\pl_\zeta\calR^*$ is the
(possibly multi-valued) convex subdifferential. The {\smaller GENERIC} evolution equation
then reads 
\[
\plt q = \bbJ(q) \rmD\calE(q) + \bbK(q) \rmD \calS(q)\,,
\]
and the second non-interaction condition is replaced by 
\[
\calR^*(q,\lambda \rmD\calE(q))\equiv 0 \quad \text{ for all }\lambda \in \R\,. 
\]
By convexity, the latter condition implies $\calR^*(q,\zeta{+}\lambda
\rmD\calE(q)) = \calR^*(q,\zeta)$ for all $(q,\zeta) \in \rmT^*\bfQ$ and $\lambda
\in \R$. Again, energy conservation and entropy increase follow.

\subsection{Arbitrary thermal variable and stress tensors}\label{su:ChoiceStress}

Using the generalization of \cite{Miel11FTDM} (see also the recent usages in
\cite{BetSch19EMEC,SchBet21SPST}), we work with a general scalar thermodynamic
variable $w$ that can denote either the temperature $\theta$, the internal
energy $e$, the entropy $s$, or other. The only restriction is that both
functions $e=E(\Fe,\alpha,\beta,w)$ and $s=S(\Fe,\alpha,\beta,w)$ satisfy the relation
\begin{equation}
  \label{eq:TemperaDef}
\theta=\Theta(\Fe,\alpha,\beta,w)=
\frac{\pl_w E(\Fe,\alpha,\beta,w)}{\pl_wS(\Fe,\alpha,\beta,w)}>0\,,
\end{equation}
where $\theta$ is the absolute temperature.

Since the derivatives of the free energy with respect to $\Fe$, $\alpha$, and
$\beta$ at constant temperature play a crutial role as thermodynamical driving
forces, we recall from \cite[Eqn.\,(2.13)]{Miel11FTDM} the important relations
\begin{subequations}
\label{eq:Deriv.psi}
\begin{align}
\label{eq:Deriv.psi.a}
&\psi'_{\Fe} (\Fe,\alpha,\beta,\theta)\big|_{\theta=\Theta(\Fe,\alpha,\beta,w)} =
E'_{\Fe}(\Fe,\alpha,\beta,w) - \Theta(\Fe,\alpha,\beta,w) S'_{\Fe}(\Fe,\alpha,\beta,w)\  
\text{ and} \\
\label{eq:Deriv.psi.b}&
\psi'_\zeta (\Fe,\alpha,\beta,\theta)\big|_{\theta=\Theta(\Fe,\alpha,\beta,w)} =
E'_\zeta(\Fe,\alpha,\beta,w) - \Theta(\Fe,\alpha,\beta,w) S'_\zeta(\Fe,\alpha,\beta,w)
\end{align}
\end{subequations}
for $\zeta=\alpha,\beta$.
For general choices of the thermal variable $w$, the free energy is
always given by $\psi(\Fe,\zeta,w)=E(\Fe,\zeta,w)-\Theta(\Fe,\zeta,w) S(\Fe,\zeta,w)$.
Note that, in view of \eqref{eq:Deriv.psi.a}, the first Piola-Kirchhoff
tensor can be calculated by the same formula $E'_{\Fe}-\Theta S'_{\Fe}$
independent of the choice of $w$, whereas the formula $\psi'_{\Fe}$ gives the first
Piola-Kirchhof tensor only for the choice $w=\theta$. Similarly, \eqref{eq:Deriv.psi.b}
says that the chemical potentials can always be calculated by $\mu^\zeta =E'_\zeta-\Theta
S'_\zeta$, but the formula $\mu^\zeta=\psi'_\zeta$ only holds for $w=\theta$. 

Our densities $E$ and $S$ are defined with respect to the actual Eulerian
volume $\d\xx$. The first Piola-Kirchhoff tensor $\TT$ is the obtained by the
derivative of the free energy (at fixed $\theta$) when taking with respect to
the material (i.e.\ Lagrangian) volume measure $\d\XX$. Using
$\d\xx=\det\Fe\d\XX$
(recall $\det \bfF_\rmp=1$) we have $\TT = \big(\det\Fe\,\psi\big)'_{\Fe}$.
The actual stress tensor in the Eulerian setting is the 
Cauchy stress tensor $\bfSigma_\mathrm{Cauchy}$ which is related to $\TT$ via
$\bfSigma_\mathrm{Cauchy}= (1/\!\det\Fe) \TT \Fe^\top$, see e.g.\
\cite[Eqn.\,(48.2)]{GuFrAn10MTC}.  Thus, exploiting $(\det\Fe)'_{\Fe} =
\mathrm{Cof}\,\Fe = (\det \Fe) \,\Fe^{-\top}$ and \eqref{eq:Deriv.psi.a}
provides us with the formula 
\begin{equation}
  \label{eq:GEN.CauchyStress}
  \bfSigma_\mathrm{Cauchy} = \frac1{\det \Fe\!}\Big( \big(E \det \Fe)'_{\Fe}\!-
\Theta \big(S\det\Fe)'_{\Fe}\Big)\Fe^\top\!=
\big(E'_{\Fe}{-}\Theta\,S'_{\Fe}\big)\Fe^\top\!+\big(E{-}\Theta\,S\big)\,\bbI.\
\end{equation}

\subsection{The Hamiltonian part of Eulerian thermoelastoplasticity}
\label{su:GEN.Hamil}
We follow ideas from \cite{HutTer08FAEM} for the Hamiltonian part of
thermoelasticity in the Eulerian setting based on the state variables
$(\pp,\Fe,\alpha,\beta,\Fe,\theta)$, see also \cite{PaPeKl20HCM} for a related 
form involving $\varrho$ as an additional state variable.

Using that $\varrho=\rho_\text{\sc r}/\!\det\Fe$, the total energy and total
entropy can be written as 
\[
\calE(\pp,\Fe,\alpha,\beta,w)=\!\int_\varOmega\!\Big( \frac{|\pp|^2}{2\varrho} +
E(\Fe,\alpha,\beta,w) \Big)\d\xx \ \text{ and } \  
\calS(\Fe,\alpha,\beta,w)=\!\int_\varOmega\! S(\Fe,\alpha,\beta,w)\d\xx.
\]
For generality we have introduced two scalar variables $\alpha$ and $\beta$,
which are assumed to be intensive and extensive, respectively. See
\cite{ZaPeTh23GFRF} for the importance of treating intensive and extensive
variables accordingly. Here, $E$ and
$S$ are the energy and entropy densities with respect to the actual Eulerian
volume. 

Taking into account $\pp=\varrho\vv$ with $\varrho=\varrho(\Fe)$
from \eq{rho} so that $\varrho'_{\Fe}=-\varrho\Fe^{-\top}$, the differentials of
$\calE$ and $\calS$ take the form 
\begin{equation}
  \label{eq:DE.DS}
  \rmD\calE(\pp,\Fe, \alpha, \beta,w)=\bma{c}\hspace{-2em}\vv
  \\[-.3em]\hspace{-.5em}E'_{\Fe}{+}\dfrac{\varrho|\vv|^2}2\Fe^{-\top}\hspace{-.5em}
  \\[-.3em]\hspace{-2em} E'_\alpha\\\hspace{-2em} E'_\beta \\\hspace{-2em} E'_w  \ema
\ \ \text{ and }\ \
\rmD\calS(\pp,\Fe, \alpha, \beta,w)=\bma{c}0\\ S'_{\Fe}\\ S'_\alpha
\\S'_\beta\\ S'_w\ema. 
\end{equation}

In {\smaller GENERIC}, the Hamiltonian (or reversible) part of the evolution is given in
the form 
\[
\plt q = \bbJ(q) \rmD\calE(q), 
\]
where $\bbJ$ satisfies three conditions:
\begin{subequations}
\label{eq:Poisson}
\begin{align}
\label{eq:Poisson.i}
&\text{skew symmetry:}\quad \bbJ(q)^*=- \bbJ(q)\,;
\\
\label{eq:Poisson.ii}
&\text{Jacobi's identity:}\quad\big\langle \zeta_1,
\rmD\bbJ(q)[\bbJ(q)\zeta_2]\zeta_3\big\rangle+\text{cycl.perm} \equiv 0\,;
\\
\label{eq:Poisson.iii}
&\text{first non-interaction condition:}\quad \bbJ(q)\rmD\calS(q)\equiv 0\,.  
\end{align}
\end{subequations}
We construct a suitable $\bbJ$ with the block structure
\begin{equation}
  \label{eq:bbJ.BlockStr}
\bbJ(q)= \bma{ccccc} 
\bbJ^{\pp\pp} &\bbJ^{\pp\Fe} & \bbJ^{\pp\alpha} & \bbJ^{\pp\beta} &\bbJ^{\pp w} \\
\bbJ^{\Fe\pp} &0  &0 & 0 &0 \\
\bbJ^{\alpha\pp} &0  &0 &0 &0 \\
\bbJ^{\beta\pp} &0  &0 &0 &0 \\
\bbJ^{w\pp} &0  &0 & 0 &0 
\ema
\end{equation}
We set 
\begin{align*}
\hspace*{3em}&
\bbJ^{\pp\pp}(q)\bfzeta
=  -\,\DIV(\pp {\otimes}\bfzeta)- (\nabla \bfzeta)^\top \pp , 
&& \bbJ^{\Fe\pp}(q)\bfzeta= -\,(\bfzeta{\cdot}\nabla ) \Fe +
    (\nabla\bfzeta)\Fe,\hspace*{3em} 
\\
&\bbJ^{\alpha\pp}(q)\bfzeta= -\,\bfzeta{\cdot}\nabla \alpha, 
&& \bbJ^{\beta\pp}(q)\bfzeta= -\,\DIV(\beta\bfzeta).
\end{align*}
Here $\bbJ^{\pp\pp}( q)$ is different form the canonical co-symplectic
structure for the incompressible Euler equation (cf.\
\cite{HutTer08FAEM,PaPeKl20HCM}); for the compressible case we follow
\cite{ZaPeTh23GFRF}. The operator $\bbJ^{\Fe\pp}$ is chosen for the
transport of the tensor $\Fe$, giving $\plt{}\Fe+(\vv{\cdot}\nabla)\Fe=(\nabla \vv)\Fe$.
In the lower line, $\bbJ^{\alpha\pp}$ and $\bbJ^{\beta\pp}$ give the simple transport of an
intensive and an extensive scalar, respectively. The operator $\bbJ^{w\pp}$ for
the transport of $w$ will be more complicated, as it has to be compatible with
the transport of the extensive variable $s=S(\Fe,\alpha,\beta,w)$, where $w$
may be neither intensive (like $w=\theta$) or extensive (like for
$w\in \{e,s\}$). The proof of the validity of Jacobi's identity in Theorem
\ref{th:PoissonStruct} relies heavily on the fact that the operators
$\bbJ^{\pp\pp}$, $\bbJ^{\Fe\pp}$, $\bbJ^{\alpha\pp}$, and $\bbJ^{\beta\pp}$ are given
by classical Lie derivatives of tensors in the direction of the vector field
$\vv=\bfzeta$. 

In view of the desired skew symmetry \eqref{eq:Poisson.i} we define 
\[
\bbJ^{\pp\Fe}( q)\bfXi= \nabla \Fe {:}\bfXi + \DIV\big(\bfXi
\Fe^\top\big), \qquad \bbJ^{\pp\alpha}( q)a= a \nabla \alpha, \qquad 
\bbJ^{\pp\beta}( q)b= -\beta \nabla b\,.  
\]

To find $\bbJ^{\pp w}$, we use the first non-interaction condition
\eqref{eq:Poisson.iii} and observe that because of $\rmD_\pp\calS( q)=0$ we
only have to satisfy the relation
\[
\bbJ^{\pp\Fe}( q)S'_{\Fe}  +
\bbJ^{\pp\alpha}( q) S'_\alpha  +
\bbJ^{\pp\beta}( q) S'_\beta  +
\bbJ^{\pp w}( q) S'_w \equiv 0\,.  
\]
Obviously, this is satisfied by the choice 
\[
\bbJ^{\pp w}( q)\omega = - S \nabla \Big(\frac{\omega}{S'_w}\Big) 
 -\bbJ^{\pp\Fe}\Big(\frac\omega{S'_w} S'_{\Fe}\Big) 
 -\bbJ^{\pp\alpha}\Big(\frac\omega{S'_w} S'_\alpha\Big) 
 -\bbJ^{\pp\beta}\Big(\frac\omega{S'_w} S'_\beta\Big) .
\]
Finally, defining $\bbJ^{w\pp}( q)= -\bbJ^{w\pp}( q)^*$ we find
\begin{align*}
\bbJ^{w\pp}( q) \bfzeta&= -\frac1{S'_w}\DIV(S\bfzeta)
+\frac1{S'_w}S'_{\Fe}{:}\,\bbJ^{\Fe \pp} \bfzeta 
+\frac1{S'_w}S'_\alpha \bbJ^{\alpha \pp} \bfzeta 
+\frac1{S'_w}S'_\beta \bbJ^{\beta \pp} \bfzeta \\
&= -\frac1{S'_w}\Big( \DIV(S\bfzeta) + S'_{\Fe}{:}\big((\bfzeta{\cdot}\nabla ) \Fe {-}
    (\nabla\bfzeta)\Fe\big) + S'_\alpha \bfzeta{\cdot}\nabla \alpha 
  + S'_\beta \DIV(\beta \bfzeta)\Big) .
\end{align*}
Thus, we have $\bbJ^*=-\bbJ$ and the full skew symmetry \eqref{eq:Poisson.i} is
established.  

Moreover, assuming purely Hamiltonian flow with
$\plt{}(\Fe,\alpha,\beta,w) = (\bbJ^{\Fe \pp},\bbJ^{\alpha \pp}, \bbJ^{\beta
  \pp},\bbJ^{w\pp})\vv$, we find 
\begin{equation}
  \label{eq:S.transported}
  \plt{} \big( S(\Fe,\alpha,\beta,w)\big) = S'_{\Fe}{:}\plt{\Fe} + S'_\alpha
\plt\alpha + S'_\beta \plt\beta + S'_w \plt w = -\DIV \big(
S(\Fe,\alpha,\beta,w)\vv\big),
\end{equation}
which shows that the entropy density is transported as an extensive variable.   

\begin{theorem}[Poisson structure.]\label{th:PoissonStruct}
The operator $\bbJ$ defined above is a Poisson structure satisfying the
conditions \eqref{eq:Poisson}. 
\end{theorem}

For the readers convenience we give an explicit and self-contained proof of the
validity of the Jacobi identity \eqref{eq:GEN.Jacobi}, however proofs for
various restricted operators $\bbJ$ exist in the literature, see e.g.\
\cite{Otti05BET, HutTer08FAEM, PaPeKl20HCM, ZaPeTh23GFRF}. Our proof will be
based on two observations: (i) $\bbJ$ has the block
structure \eqref{eq:bbJ.BlockStr}, which will be analyzed in Appendix
\ref{app:PoissBlock}, and (ii) that the components of $\bbJ$ are given in terms
of Lie derivatives of tensors with respect to the underlying velocity field
$\vv$.

For a general vector field $\ww$ and a tensor field $\TT$, the Lie derivative
is defined by taking the derivative of $\TT$ along the flow of the $\ww$.
The important property of Lie derivatives is the commutator relation 
\begin{equation}
  \label{eq:LieCommutator}
  \mfL_\vv\big( \mfL_\ww \TT\big) - \mfL_\ww \big(
\mfL_\vv \TT\big) = \mfL_{[[\vv,\ww]]} \TT,
\end{equation}
where the commutator between vector fields is given by 
\[
[[\vv,\ww]]:= \mfL_\vv \ww = - \mfL_\ww \vv = (\vv{\cdot}\nabla)\ww -
  (\ww{\cdot}\nabla) \vv= (\nabla \ww) \vv -(\nabla \vv) \ww .
\]
The identity \eqref{eq:LieCommutator} cannot be found easily in the literature,
but it is an easy consequence of its validity for functions, vectors, and
co-vectors ($1$-forms) and of the well-known derivation rule $\mfL_\vv (\TT{\otimes}\SS)
= (\mfL_\vv \TT){\otimes}\SS + \TT {\otimes}(\mfL_\vv \SS)$ by doing induction over
the rank of the tensors. 

\begin{proposition}[Lie derivatives]
\label{pr:LieDeriv}
We have the following identities 
\begin{subequations}
  \label{eq:LieDeriv.bbJ}
\begin{align}
  \label{eq:Lie.bbJ.a}  
 -\bbJ^{\pp\pp}(\qq) \vv& = \mfL_\vv^{(d-1)\text{\rm-fo}}\,\qq =
   \DIV(\qq{\otimes} \vv) + (\nabla \vv)^\top \qq , 
\\
  \label{eq:Lie.bbJ.b} 
 -\bbJ^{\Fe\pp}(\GG) \vv& = \mfL^{\mathrm{vec}}_\vv \,\GG = (\vv{\cdot}
 \nabla) \GG - (\nabla\vv)\GG ,
\\
  \label{eq:Lie.bbJ.c}
 - \bbJ^{\alpha\pp}(a) &= \mfL^{0\text{\rm-fo}}_\vv \, a = (\vv{\cdot} \nabla) a = \vv
  \cdot \nabla a,
\\
  \label{eq:Lie.bbJ.d}
   - \bbJ^{\beta\pp}(b)&=\mfL^{d\text{\rm-fo}}_\vv \,b = \DIV(b \vv), 
\end{align}
where $\mfL^{n\text{\rm-fo}}_\vv$ denotes the Lie derivatives of an
$n$-form, where intensive variables are $0$-forms (simple functions) and
extensive variables are $d$-forms (densities of volume forms with respect to
the Lebesgue measure).  
\end{subequations}
\end{proposition} 
\noindent{\it Proof.} The relations in \eqref{eq:Lie.bbJ.c} and
\eqref{eq:Lie.bbJ.d} are in all textbooks on tensor calculus, see e.g.\
\cite[Sec.\,4.3]{MarRat99IMS}. The same holds for \eqref{eq:Lie.bbJ.b}, if we
observe that $\GG \mapsto \bbJ^{\Fe\pp}(\GG) \vv$ acts on the columns of $\GG$,
i.e.\ we interpret $\GG$ as a collection of column vectors.  

To obtain \eqref{eq:Lie.bbJ.a} we can use the duality between
$(d{-}1)$-forms and vectors. With $ \mfL^{1\text{\rm-fo}}_\vv\bfxi=
(\vv{\cdot}\nabla) \bfxi + (\nabla \vv)^\top \bfxi$ we find 
\begin{align}\nonumber
\int_{\varOmega}\ww\cdot \mfL_\vv^{(d-1)\text{\rm-fo}}(\qq)\,\d\xx
&:= -\int_{\varOmega}\qq \cdot \mfL^{\text{\rm vec}}_\vv \ww \,\d\xx
= \int_{\varOmega} \qq\cdot\big(
(\vv{\cdot}\nabla) \ww-(\nabla \vv) \ww\big) \,\d\xx
\\[-.3em]\label{eq:mfLvec.mfLd-1}
&  = \int_{\varOmega} \ww\cdot
\big( \DIV(\qq{\otimes}\vv)+(\nabla \vv)^\top \qq\big) \,\d\xx\,,
\end{align}
which is the desired result. 
\mbox{}\hfill$\Box$

\medskip

We are now ready to complete the

\noindent{\it Proof of Theorem \ref{th:PoissonStruct}.}
The conditions  \eqref{eq:Poisson.i} and \eqref{eq:Poisson.iii} are satisfied
by construction.

To show the Jacobi identity \eqref{eq:Poisson.ii}, there are two classical
ways: (i) one simply uses the definition and evaluates the tri-linear form
defined in \eqref{eq:Poisson.ii} or (ii) one uses the invariance of the Jacobi
identity under the coordinate transformation. 

We choose the latter one and consider the new variables
$\ol q =(\pp,\Fe,\alpha,\beta,s)$ with $s=S(\Fe,\alpha,\beta,w)$.  By
\eqref{eq:TemperaDef} we know that this mapping is invertible to obtain
$w=W(\Fe,\alpha,\beta,s)$ back again.  The new form of the energy density is
$\ol E(\pp,\Fe,\alpha,\beta,s)=
E\big(\pp,\Fe,\alpha,\beta,W(\Fe,\alpha,\beta,s)\big) $ and the entropy
density is $\ol S(\pp,\Fe,\alpha,\beta,s)=s$. In particular, using
$\ol S'_{\Fe}=0$, $\ol S'_\alpha=S'_\beta=0$, and $\ol S'_s=1$, the operator
$\bbJ$ transforms into
\[
\bbJ(q)= \bma{ccccc} 
- \mfL^{(d-1)\text{\rm-fo}}_\Box (\pp)&\bbJ^{\pp\Fe}(\Fe) & \Box\nabla
      \alpha &- \beta\nabla\Box & -s \nabla \Box\\
- \mfL^\text{vec}_\Box(\Fe) &0  &0 & 0 &0 \\
-\mfL^{0\text{\rm-fo}}_\Box ( \alpha) &0  &0 &0 &0 \\
-\mfL^{d\text{\rm-fo}}_\Box ( \beta) &0  &0 &0 &0 \\
-\mfL^{d\text{\rm-fo}}_\Box ( s) &0  &0 & 0 &0 
\ema,
\]
where we already inserted the results from Proposition \ref{pr:LieDeriv}. 

Finally, we apply Proposition \ref{pr:Jacobi:Ident} with $n\in \{1,\ldots,N\}$
replaced by $\aa\in \{\pp,\Fe,\alpha,\beta,s\}$, i.e.\ $\pp$ plays the special role of
$n=1$ in the block structure. We first observe that all the operators
$\aa\mapsto \bbJ^{\aa\pp}(\aa)$ are linear, such that
\[
\rmD\bbJ^{\aa\pp}(\aa)[\bbJ^{\aa\pp}(\aa)\vv] \ww
= \bbJ^{\aa\pp}\big(\bbJ^{\aa\pp}(\aa)\vv\big)\ww  
= \mfL^\aa_\ww \big(  \mfL^\aa_\vv \aa\big) ,
\]
where $\mfL^\aa_\vv$ stands for $\mfL^{(d-1)\text{\rm-fo}}_\vv$, $
\mfL^\text{vec}_\vv$, $\mfL^{0\text{\rm-fo}}_\vv$, $ \mfL^{d\text{\rm-fo}}_\vv$,
and $\mfL^{d\text{\rm-fo}}_\vv$, respectively. 

To establish the Jacobi identity for $\bbJ^{\pp\pp}$ we use that
$\mfL^\text{vec}_\vv \ww=[[\vv,\ww]]$ satisfies it (see
e.g.\,\cite[p.\,143]{MarRat99IMS}) and dualize:
\begin{align*}  
\big\langle \vv_1,\bbJ^{\pp\pp}\big(\bbJ^{\pp\pp}(\pp)\vv_2\big)
    \vv_3\big\rangle + \text{cycl.\,perm.} 
&= \big\langle \vv_1,\mfL^\pp_{\vv_2} \mfL^\pp_{\vv_3}\pp \big\rangle +
  \text{cycl.\,perm.} 
\\
= \big\langle \mfL^\text{vec}_{\vv_3}\mfL^\text{vec}_{\vv_2} \vv_1, \pp
   \big\rangle + \text{cycl.\,perm.} 
& = \big\langle [[\,\vv_3,\, [[\vv_2, \vv_1]]\:]]+ \text{cycl.\,perm.} , \pp
   \big\rangle = \langle 0,\pp\rangle = 0.
\end{align*}
Thus, \eqref{eq:Conds.JI.A} and \eqref{eq:Conds.JI.B} are established.
To obtain \eqref{eq:Conds.JI.C} for $\aa\in \{\Fe,\alpha,\beta,s\}$ we use the
commutator property 
\eqref{eq:LieCommutator} for Lie derivatives. 
\begin{align*}
&\big\langle \zeta_\aa, \bbJ^{\aa\pp}(\bbJ^{\aa\pp}(\aa)\vv)\ww {-} 
   \bbJ^{\aa\pp}(\bbJ^{\aa\pp}(\aa)\ww)\vv \big\rangle
 = \big\langle \zeta_\aa, \mfL^\aa_\ww \mfL^\aa_\vv \aa {-} \mfL^\aa_\vv
 \mfL^\aa_\ww  \aa\big\rangle  
\\
&\ \,\overset{\text{\eqref{eq:LieCommutator}}}= \big\langle \zeta_\aa,
\mfL^\aa_{[[\ww,\vv]]} \aa \big\rangle_{X_\aa} = -\big\langle \zeta_\aa,
-\bbJ^{\aa \pp}(a) [[\vv,\ww]] \big\rangle_{X_\aa}= -\big\langle [[\vv,\ww]],
\bbJ^{\pp\aa}(\aa)\zeta_\aa \big\rangle_{X_\pp}  \\
&\ \ \ = -\big\langle \mfL^\text{vec}_\vv \ww, \bbJ^{\pp\aa}(\aa)\zeta_\aa
\big\rangle_{X_\pp} \overset{\text{\eqref{eq:mfLvec.mfLd-1}}}=
\big\langle \ww,\mfL^\text{(d-1)\text{\rm-fo}}_\vv  \bbJ^{\pp\aa}(\aa)\zeta_\aa
\big\rangle_{X_\pp}  = \big\langle \ww, \bbJ^{\pp\pp} \big(
\bbJ^{\pp\aa}(\aa)\zeta_\aa\big) \vv \big\rangle\,.
\end{align*}
Thus, \eqref{eq:Conds.JI.C} is established as well and the proof is complete.
$\hfill\Box$

\medskip

We now discuss the terms arising from of the Hamiltonian part
of the dynamics, namely $\plt q= \bfV_\text{Ham}(q) = \bbJ(q)\rmD\calE(q)$. 
Using \eqref{eq:DE.DS} we obtain the following system of equations
\begin{equation}
  \label{eq:HamPart}
\bfV_\text{Ham}(q)=\bbJ(q)\rmD\calE(q)
 = \bma{c} -\DIV\big(\varrho\vv{\otimes}\vv\big) + \DIV\big(
 \bfSigma_\text{Cauchy}- p^{(\beta)}\bbI\big)\\
-\vv{\cdot}\nabla \Fe + (\nabla\vv)\,\Fe \\ 
-\vv{\cdot}\nabla \alpha\\ -\DIV(\beta \vv)
\\[-.0em]-\dfrac1{S'_w}\DIV(S\vv)
-\dfrac1{S'_w}\Big(\displaystyle{ S'_{\Fe}{:}\plt{\Fe}  
+S'_\alpha \plt\alpha + S'_\beta \plt\beta}\Big)\ema\,,
\end{equation}
where the Cauchy stress tensor $\bfSigma_\text{Cauchy}$ is defined via the
free-energy density $\psi=E-\Theta S$ as follows
\[
\bfSigma_\text{Cauchy} = \wt\bfSigma  \Fe^\top\!+ \psi\,\bbI
\quad \text{with }\ \wt\bfSigma = E'_{\Fe}(\Fe,\alpha,\beta,w)-\Theta(\Fe,\alpha,\beta,w)  S'_{\Fe}(\Fe,\alpha,\beta,w)\,,
\] 
see Section \ref{su:ChoiceStress}. Moreover, the
extensive variable $\beta$ generates the additional pressure 
\[
p^{(\beta)}= \beta \,(E'_\beta - \Theta S'_\beta)\,,
\]
see also \cite[Eqn.\,(2.15e)]{PaPeKl20HCM}.  

All terms in \eqref{eq:HamPart} are clear except for the first one involving
the Cauchy stress tensor. To obtain the given compact form, we start from the
definition
\begin{align*}
\bfV^\pp_\text{Ham}(q)&= \bbJ^{\pp\pp}\rmD_\pp\calE +
\bbJ^{\pp\Fe}\rmD_{\Fe}\calE+ \bbJ^{\pp\alpha}\rmD_\alpha \calE+
  \bbJ^{\pp\beta} \rmD_\beta \calE+ \bbJ^{\pp w}\rmD_w\calE
\\
&= \bbJ^{\pp\pp}(\pp)\vv +
\bbJ^{\pp\Fe}(\Fe)\Big(E'_{\Fe}+\frac{\varrho|\vv|^2}2 \Fe^{-\top}\Big) +
E'_\alpha \nabla \alpha - \beta \nabla(E'_\beta) \\
&\quad -S\nabla \Big(\frac{E'_w}{S'_w}\Big)
-\bbJ^{\pp\Fe}(\Fe)\Big(\frac{E'_w}{S'_w} S'_{\Fe}\Big) -
\Big(\frac{E'_w}{S'_w} S'_\alpha\Big) \nabla \alpha-\beta\nabla\Big(\frac{E'_w}{S'_w} S'_\beta\Big)\,. 
\end{align*}
For the terms involving $\vv$, we use $\pp=\varrho\vv$ and find an
important cancellation (see also \cite[Eqn.\,(4.6)]{ZaPeTh23GFRF}): 
\begin{align*}
&\bbJ^{\pp\pp}(\pp)\vv 
  {+} \bbJ^{\pp\Fe}(\Fe)\Big(\frac{\varrho|\vv|^2\!\!}2\Fe^{-\top}\Big)
= - \DIV(\varrho\vv{\otimes}\vv) 
 {-}  (\nabla\vv)^\top\! (\varrho\vv) 
 {+} \frac{\varrho|\vv|^2\!\!}2 \Fe^{-\top} {:} \nabla \Fe 
 {+} \DIV\Big(\frac{\varrho|\vv|^2\!\!}2\,\bbI\Big) \\
&\qquad =  - \DIV(\varrho\vv{\otimes}\vv) 
  - \varrho \,  \nabla \Big(\frac{|\vv|^2}2\Big) 
   + \frac{\varrho|\vv|^2}{2\det\Fe} \nabla \det\Fe 
   +  \nabla\Big( \varrho \frac{|\vv|^2}2 \Big) 
 =  - \DIV(\varrho\vv{\otimes}\vv),
\end{align*}
because $\varrho=\rho_\text{\sc r}/\!\det\Fe$ implies
$\nabla\varrho= -(\varrho/\!\det\Fe)\nabla\det\Fe$; here we rely on that
$\rho_\text{\sc r}$ is assumed constant in space.  

Setting $\wt\bfSigma=E'_{\Fe}{-}\Theta S'_{\Fe}$ and using $\Theta = E'_w/S'_w$
(note that $\psi'_{\Fe}=\wt\bfSigma{-}S \Theta'_{\Fe} \neq \wt\bfSigma$) and
$\psi=E-\Theta S$, we can also simply the other 
terms and find 
\begin{align*}
  \bfV^\pp_\text{Ham}(q)&=-\,\DIV(\varrho\vv{\otimes}\vv) + \bbJ^{\pp\Fe}(\Fe)
  \wt\bfSigma  +(\psi'_\alpha{+}S\Theta'_\alpha) \nabla \alpha
  -  \beta\nabla (\psi'_\beta{+}S\Theta'_\beta) - S \nabla \Theta
  \\
  &=-\,\DIV(\varrho\vv{\otimes}\vv) + \bbJ^{\pp\Fe}(\Fe) \wt\bfSigma +
  \psi'_\alpha\nabla \alpha - \beta \nabla \psi'_\beta\\
  &\qquad +S\Theta'_\alpha \nabla \alpha - \beta\nabla(S\Theta'_\beta) -
  S\big(\Theta'_{\Fe}{:}\nabla \Fe{+} \Theta'_\alpha\nabla
  \alpha{+}\Theta'_\beta\nabla \beta + \Theta'_w \nabla w \big)\\
  &= -\,\DIV(\varrho\vv{\otimes}\vv) + \bbJ^{\pp\Fe}(\Fe)  \wt\bfSigma
   -S \Theta'_{\Fe}{:}\nabla \Fe + \psi'_\alpha\nabla\alpha 
  -\beta \nabla\psi'_\beta -\nabla(S\beta \Theta'_\beta)+\psi'_w \nabla w\,,
\end{align*}
where, for the last term, we used the definition of $\psi$ and $\Theta$ to
find $\psi'_w=-S\Theta'_w$. 
On the other hand we have 
\begin{align*}
&\DIV\big(\wt\bfSigma  \Fe^\top+\psi\,\bbI\big)
=\bbJ^{\pp\Fe}(\Fe)\wt\bfSigma -\wt\bfSigma{:}\nabla \Fe + \nabla \psi\\
&\quad =\bbJ^{\pp\Fe}(\Fe)\wt\bfSigma - E'_{\Fe}{:}\nabla \Fe + \Theta
S'_{\Fe}{:}\nabla \Fe +(E{-}\Theta S)'_{\Fe}{:}\nabla \Fe  +\psi'_\alpha\nabla
\alpha + \psi'_\beta \nabla \beta+ \psi'_w \nabla w
\\
&\quad =\bbJ^{\pp\Fe}(\Fe)\wt\bfSigma -S\Theta'_{\Fe}{:}\nabla \Fe  +\psi'_\alpha\nabla
\alpha + \psi'_\beta \nabla \beta+ \psi'_w \nabla w\,.
\end{align*}
Combining the last two relations and exploiting $\psi'_\beta +
S\theta'_\beta=E'_\beta-\Theta S'_\theta$  yields 
\begin{equation}
  \label{eq:bfp.EqnSimplif}
  \bfV^\pp_\text{Ham}(q)=- \DIV(\varrho\vv{\otimes}\vv)+ \DIV\Big( \wt\bfSigma
  \Fe^\top + \big(\psi-\beta(E'_\beta{-}\Theta S'_\beta)\big)\,\bbI \Big),
\end{equation}
which shows the additional pressure correction $\beta(E'_\beta{-}\Theta
S'_\beta)$ for the extensive variable $\beta$. 

With this the form of $ \bfV_\mathrm{Ham}$ given in \eqref{eq:HamPart} is established. 
It can be checked easily, that when omitting $\alpha$ and $\beta$ and choosing
$\theta=w=\Theta(\Fe,w)$ we exactly obtain the equations derived in
\cite[Sec.\,3]{HutTer08FAEM}.

\subsection{The dissipative part of Eulerian thermoelastoplasticity}\label{su:GEN.diss}

As explained in \cite[Sec.\,2.3.2]{Otti05BET} (following \cite{Edwa98ASDG}) and
\cite[Sec.\,4.3]{Miel11FTDM}, suitable nonlinear dissipation potentials $\calR$
or linear Onsager operators $\bbK$ are constructed by collecting the building
blocks of the dissipative mechanics and then combining them
with a nontrivial operator $N_\calE$ in the form
\[
\calR^*(q,\bfzeta)=\calR^*_\mathrm{simple}(q,N_\calE(q)^*\bfzeta) \quad
\text{or} \quad \bbK(q)=N_\calE(q)\bbK_\mathrm{simple}(q)N_\calE(q)^*.
\]
In our model we can have five different dissipative processes:

(A) viscoelastic dissipation induced by $\Dv=\frac12\big(\nabla
\vv{+}(\nabla\vv)^\top \big)$,

(B) inelastic (or plastic) dissipation induced by $\Lp =
\Fe^{-1}\big((\nabla\vv)\,\Fe{-}\plt{}\Fe {-}(\vv{\cdot}\nabla)\Fe\big)$, 

(C) diffusion and growth/decay for the intensive variable $\alpha$,

(D) diffusion for the extensive variable $\beta$,

(E) heat flow induced by $\nabla (1/\theta)$. 

\noindent
Thus, we can find a suitable dual dissipation potential in the additive form 
\[
 \calR^*_\mathrm{simple} = \calR^*_\rmA + \calR^*_\rmB + \calR^*_\rmC +
 \calR^*_\rmD + \calR^*_\rmE. 
\]
However, we emphasize that this simplistic assumption is by far not
necessary. Of course, it is possible to construct much more general
thermodynamically consistent models where there is a strong interaction
of the different dissipation mechanics (e.g.\ like cross diffusion). Nevertheless, we
will restrict our approach to the case of a simple block structure. 

The advantage of using an operator $N_\calE$ is three-fold. First, it is used
guarantee the second non-interaction condition by asking
\begin{equation}
  \label{eq:NE.2ndNIC}
    N_\calE(q)^* \rmD\calE(q)=(0,...,0,1)^\top \quad \text{and} \quad
  \calR^*_\mathrm{simple}\big(q,(0,...,0,\lambda)^\top\big)=0\ \text{ for all
  }\  \lambda = \in \R.
\end{equation}
Here $\lambda\in \R$ stand for the constant (reciprocal of the) 
temperature $1/\theta$, which does not generate any dissipation.

The second advantage is the fundamental observation in
\cite[Sec.\,4.3]{Miel11FTDM} that now the dissipative
(a.k.a.\ irreversible) driving forces are now given by  
\[
\bfeta= N_\calE(q)^*\rmD\calS(q)\,,
\]
which contains already information on $\calE$ in a specific way because of
\eqref{eq:NE.2ndNIC}. 

Finally, the operator $N_\calE$ acting from the left on
$\pl\calR^*_\mathrm{simple}$ will encoded the dissipative terms that generate the energy
conservation, i.e.\ the terms that contributing to the entropy
production. 

For our special application, we recall the special form of $\rmD\calE$ and $\rmD\calS$
given in \eqref{eq:DE.DS} and construct an operator $N_\calE^*: \R^d\ti
\R^{d\ti d} \ti \R^3\to \R^{d\ti d}_\mathrm{sym}\ti \R^{d\ti d} \ti \R^3$ as follows:
\[
N_\calE(q)^*=\bma{ccccc}\Dv(\Box)&0&0&0& \frac{\ds-\Box}{\ds E'_w}\Dv\\[0.8em]
\!\!\! -\dfrac\varrho2 (\vv{\cdot}\Box )\bbI\!\!  &\:\Fe^\top&0&0&
 \frac{\ds-\Box}{\ds E'_w} \Fe^\top E'_{\Fe}\\[0.7em]
 0&0&\:1\:&0&-E'_\alpha/E'_w\\
0&0&0&\:1\:&-E'_\beta/E'_w
\\ 
0&0 &0 & 0 & 1/E'_w \ema.
\]
For the later analysis, we introduce the vector of the reduced driving forces
$\bfeta=N_\calE(q)^*\bfzeta \in \R^{d\ti d}_\mathrm{sym}\ti \R^{d\ti d} \ti \R^3$
via
\[
\bfeta=\bma{c}\bfeta^\pp\\ \bfeta^{\Fe} \\ \eta^\alpha\\ \eta^\beta\\ \eta^w
\ema = N_\calE(q)^*\bma{c} \bfzeta^\pp\\ \bfzeta^{\Fe}\\ \zeta^\alpha \\ \zeta^\beta \\
\zeta^w\ema 
 = \bma{c}\Dv(\bfzeta^\pp)-(\zeta^w/E'_w)\vv\\  \!\! \Fe^\top\bfzeta^{\Fe} -
 \dfrac\varrho2(\vv{\cdot}\bfzeta^\pp) \bbI -(\zeta^w/E'_w)\Fe^\top E'_{\Fe} \!\! \\
  \zeta^\alpha - \zeta^w E'_\alpha/E'_w\\ 
  \zeta^\beta - \zeta^w E'_\beta/E'_w\\   \zeta^w/E'_w\ema\,.
\]

Clearly, the first relation in \eqref{eq:NE.2ndNIC} is satisfied, and
using \eq{eq:DE.DS} again we have 
\[
N_\calE(q)^*\,\rmD\calS(q) = \bma{c}-(1/\Theta) \Dv\\[0.1em]
  -(1/\Theta)\,\Fe^\top\big(E'_{\Fe}{-}\Theta S'_{\Fe}\big) \\[0.5em] 
  -(1/\Theta)(E'_\alpha{-}\Theta S'_\alpha ) \\[0.5em] 
  -(1/\Theta) (E'_\beta{-}\Theta S'_\beta )\\[0.5em]
   1/\Theta \ema\,. 
\]
The third and fourth component are given in terms of the chemical potentials
$\mu^{\alpha,\beta}$ in the form $\eta^\alpha=-\mu^\alpha/\Theta$
and $\eta^\beta=-\mu^\beta/\Theta$, because of \eqref{eq:Deriv.psi.b}.

For the adjoint operator $N_\calE: \R^{d\ti d}_\mathrm{sym}\ti
\R^{d\ti d} \ti \R^3\to \R^d\ti \R^{d\ti d} \ti \R^3$  we obtain 
\[
N_\calE(q)=\bma{ccccc} \DIV(\Box)&\!\!\! 
-\dfrac\varrho2 \mathrm{tr}(\Box)\,\vv\!\!\!\!\! &0&0&0\\
 0 &\Fe&0&0&0\\ 0&0&1&0&0\\ 0&0&0&1&0 \\ 
\frac{\ds -1}{\ds E'_w} \Dv{:}\Box& \frac{\ds -1}{\ds E'_w} 
(\Fe^\top E'_{\Fe}){:}\Box&
-E'_\alpha/E'_w & -E'_\beta/E'_w & 1/E'_w \ema .
\]
The non-diagonal entries in the last line will provide energy conservation as
well as entropy production, see Section \ref{su:EnergyEntropy}. 

The full dual dissipation potential takes the form
\[
\calR^*(q,\bfzeta)=\calR^*_\mathrm{simple}(q,N_\calE(q)^*\bfzeta)\,,
\]
and now assume that $\calR^*_\mathrm{simple}$ has a block structure
\begin{align*}
\calR^*_\mathrm{simple}(q,\bfeta) &=\calR^*_\mathrm{visc}(q,\bfeta^\pp) +
\calR^*_\mathrm{plast} (q,\bfeta^{\Fe})+ \calR^*_{\alpha}(q,\eta^\alpha) +
\calR^*_{\beta}(q,\eta^\beta)  + \calR^*_\mathrm{heat}(q,\eta^w)\,.
\end{align*}
Yet, hasten to say that this is a simplification that is not necessary at
all and, actually \eq{SMA-dissip} below would need a fully general structure.
In fact, it is one of the big advantages of the {\smaller GENERIC} framework
that it can easily handle coupling phenomena between different effects. 

Even for the these five blocks we only write the simplest forms and leave the
study of more general dissipation potentials to future work. 
\begin{subequations}\label{dissip-pot}\begin{align}
\label{dissip-pot-a}
&\calR^*_\mathrm{visc}(q,\bfeta^\pp) = \int_\varOmega \frac{\Theta}2 \bfeta^\pp{:}
\bbD_\mathrm{visc} (q)\bfeta^\pp\,\d\xx,
\quad
\calR^*_\mathrm{plast} (q,\bfeta^{\Fe}) =\!\int_\varOmega\! R_\mathrm{plast}^*
   \big(q,-\Theta \bbM \bfeta^{\Fe}\big)\,\d\xx,
\\
&\calR^*_{\alpha}(q,\eta^\alpha)=\int_\varOmega\frac12 \nabla
\eta^\alpha{\cdot}\bbA_\mathrm{diff}(q) \nabla \eta^\alpha + 
     \frac{A_\mathrm{source}(q)}{2} ( \eta^\alpha)^2\,\d\xx, 
\\&
\calR^*_{\beta}(q,\eta^\beta)=\!\int_\varOmega \frac12 \nabla
\eta^\beta{\cdot}\bbB_\mathrm{diff}(q) \nabla \eta^\beta\,\d\xx ,
\quad\
\calR^*_\mathrm{heat}(q,\eta^w) =\!\int_\varOmega \frac12 \nabla
\eta^w{\cdot}\bbK_\mathrm{heat}(q) \nabla \eta^w\,\d\xx .
\end{align}\end{subequations}

With these choices we can write down the dissipative (irreversible) part of the
evolution:
\begin{align*}
\bfV_\mathrm{irr}(q)&= 
\bma{c}\DIV\big( \bbD_\mathrm{visc}(q)\Dv +\dfrac\varrho2 \mathrm{tr}(
\Lp) \vv  \\[-.3em]  -\Fe \Lp
\\
V^\alpha_\mathrm{irr}(q):=
\DIV\Big(\bbA_\mathrm{diff} \nabla\dfrac{\mu^\alpha}\Theta\Big)
-A_\mathrm{source}\dfrac{\mu^\alpha}\Theta
\\  
V^\beta_\mathrm{irr}(q):=\DIV\Big(\bbB_\mathrm{diff}\nabla\dfrac{\mu^\beta}\Theta\Big) \ \ \ 
 \\
\!\!
 \dfrac1{E'_w} \Dv{:} \bbD_\mathrm{visc} \Dv
 + \dfrac1{E'_w}E'_{\Fe} {:} (\Fe \Lp) 
 - \dfrac{E'_\alpha}{E'_w}  V^\alpha_\mathrm{irr} 
 - \dfrac{E'_\beta}{E'_w} V^\beta_\mathrm{irr}
 - \dfrac{1}{E'_w}\DIV\!\Big(\bbK_\mathrm{heat}\nabla\dfrac1\Theta\Big)
\!\!   \ema\,,
\end{align*}
where
\[
\Lp = \theta \bbM^*\pl_{\bfeta^{\Fe}} R^*_\mathrm{plast} \big(q, \bbM(q)
\hspace{-1em}\linesunder{\Fe^\top(E'_{\Fe}{-}\Theta S'_{\Fe})}{=Mandel-stress tensor}{(symmetric)}\hspace{-1em}\big)\,.
\]

\subsection{The full {\smaller GENERIC} evolution equation}\label{su:FullGEE}

We can now assemble the whole {\smaller GENERIC} evolution equations
\[
\dot q = \bfV_\mathrm{Ham}(q) + \bfV_\mathrm{irr}(q)= \bbJ(q)\rmD\calE(q) +
\bbK(q) \rmD\calS(q). 
\]
This leads to the following system for $(\vv,\Fe,\alpha,\beta,w)$:
\begin{subequations}\label{system}\begin{align}\label{system-a}
  \plt{}(\varrho\vv) {+} \DIV\big(\varrho\vv{\otimes}\vv)
  &=\DIV\big(\bfSigma_\mathrm{Cauchy}-\beta(E'_\beta{-}\Theta S'_\beta)\bbI +
  \bbD_\mathrm{visc}(q) \Dv \big) + \frac\varrho2\,\mathrm{tr}(\Lp)\vv\,,
  \\[-.4em]\label{system-b}
  \plt{\Fe}&=-(\vv{\cdot}\nabla)\Fe + (\nabla \vv) \Fe - \Fe \Lp\,,
  \\\label{system-c}
  \plt\alpha&=-\vv{\cdot}\nabla \alpha 
  +\DIV\Big(\bbA_\mathrm{diff} \nabla\frac{\mu^\alpha}\Theta\Big)
  - A_\mathrm{source}\frac{\mu^\alpha}\Theta\,,
  \\\label{system-d}
  \plt \beta&= -\DIV(\beta\vv)
  +\DIV\Big(\bbB_\mathrm{diff} \nabla\frac{\mu^\beta}\Theta\Big)\,,
  \\\label{system-e}
  \plt w&=V^S_\mathrm{Ham} +V^E_\mathrm{irr} - \frac{\ds1}{\ds E'_w} \DIV\!\Big(
\bbK_\mathrm{heat} \nabla\frac1\Theta\Big),
\end{align}\end{subequations}
where $\varrho$ is from \eq{rho} and where we have 
\begin{subequations}\label{system-data}\begin{align}\label{system-data-a}
\bfSigma_\mathrm{Cauchy}&=\big( E'_{\Fe} {-} \Theta S'_{\Fe}\big)\Fe^\top
                        + (E{-}\Theta S) \bbI,
\\\label{system-data-b}
\Lp&= \theta \bbM^*\pl_{\bfeta^{\Fe}} R^*_\mathrm{plast} \big(q, \bbM(q)
\bfSigma_\mathrm{Mandel}\big) \quad \text{ with }\bfSigma_\mathrm{Mandel} =
\Fe^\top(E'_{\Fe}{-}\Theta S'_{\Fe}), 
\\\label{system-data-c}
\mu^\alpha&=E'_\alpha - \Theta S'_\alpha \quad \text{and} \quad 
\mu^\beta=E'_\beta - \Theta S'_\beta,
\\\label{system-data-d}
V^S_\mathrm{Ham}&= \frac1{S'_w}\Big( \! {-}\DIV(S\vv) - S'_{\Fe} {:} \plt{\Fe} -
S'_\alpha \plt \alpha - S'_\beta \plt \beta \Big) ,
\\\label{system-data-e}
V^E_\mathrm{irr}&= \frac1{E'_w} \Big(-\DIV(E\vv)+\Dv{:}\bbD_\mathrm{visc}\Dv 
    + E'_{\Fe}{:}(\Fe \Lp) - E'_\alpha \plt\alpha- E'_\beta \plt\beta\Big).
\end{align}\end{subequations}

\subsection{Energy conservation and entropy production}\label{su:EnergyEntropy}

By construction, the solutions of a {\smaller GENERIC} system automatically
satisfy the conservation of the total energy and the non-decay of the total
entropy, namely
\[
\frac{\rmd}{\rmd t} \calE(q(t)) = 0 \qquad \text{and} \qquad 
\frac{\rmd}{\rmd t} \calS(q(t)) = \big\langle \rmD
\calS(q(t)),\pl_\zeta\calR^*\big(q(t),\rmD\calS(q(t)) \big) \big\rangle \geq 0.
\]
However, in continuum mechanics it is also important to understand the local
balance laws involving the mechanical power $p_\mathrm{mech}$, the energy flux
$\bfj_\mathrm{ener}$, the entropy flux $\bfj_\mathrm{entr}$, and the entropy
production $\sigma_\mathrm{prod}$ in the form
\begin{subequations}
  \label{eq:LocalBalLaws}
\begin{align}
\label{eq:LocalEnergy}
\plt E + \DIV(E\vv) &= p_\mathrm{mech} - \DIV\bfj_\mathrm{ener}\quad \text{ and}
\\
\label{eq:LocalEntropy}
\plt S + \DIV(S\vv) &= \sigma_\mathrm{prod} -
\DIV\bfj_\mathrm{entr} \quad\text{ with }\ \sigma_\mathrm{prod}\geq 0
\text{ in }\varOmega\,. 
\end{align} 
\end{subequations}
Here $ p_\mathrm{mech}$ is due to the exchange of kinetic to the potential
(=internal) energy. For deriving these local balance laws, we take advantage of
the possibility of choosing an arbitrary thermal variable $w$. For deriving
\eqref{eq:LocalEnergy} it is advantageous to choose
$w=e=E(\Fe,\alpha,\beta,e)$, while for deriving \eqref{eq:LocalEntropy} it is
advantageous to choose $w=s=S(\Fe,\alpha,\beta,s)$.

In particular, this subsection highlights the role of the specific form of the
equation for $w$ that involves the terms $V^S_\mathrm{Ham}$ and
$ V^E_\mathrm{irr}$. These terms arise in a specific way by the construction of
the {\smaller GENERIC} formulation.

The term $V^E_\mathrm{irr}$ stems from $N_\calE$ and can be seen as a providing
the heat production through the dissipative processes. This is most elegantly
seen by using our freedom to choose the thermal variable at our will:
if we now make
\[
\text{the specific choice} \quad w=e, \qquad \text{i.e.} \quad
E(\Fe,\alpha,\beta,e)=e\,,
\]
we find $E'_{\Fe}=\bm0$, $E'_\alpha=0=E'_\beta$, $E'_e=1$ and
$S'_e=1/\Theta$. With this choice, 
the equation for the extensive variable $w=e$ takes the form  
\begin{align*}
\plt e= \Theta\big( {-}\DIV(S\vv) - S'_{\Fe}{:} \big((v{\cdot} \nabla)\Fe -
(\nabla\vv)\Fe+\Fe\Lp\big) +S'_\alpha\vv {\cdot} \nabla \alpha + S'_\beta \DIV(\beta
\vv)  \Big)\ \\
\quad   +\,\Dv{:}\bbD_\mathrm{visc}\Dv -
\DIV\!\Big(\bbK_\mathrm{heat}\nabla\frac1\Theta\Big)\,.
\end{align*}
With $\Theta\DIV(S\vv)= \DIV(e\vv)-(e{-}\Theta S)\bbI{:}\Dv + \Theta
\big(S'_{\Fe}{:}(\vv{\cdot} \nabla) \Fe +S'_\alpha \vv{\cdot}\nabla \alpha + 
S'_\beta \vv{\cdot}\nabla \beta \big)$ we see several cancellations and
arrive at 
\begin{align}\label{energy-eq}
\pdt e + \DIV(e \vv)&= \!\!\!\!\!\lineunder{\big( \bfSigma_\mathrm{Cauchy}-\beta(E'_\beta{-}\Theta
S'_\beta)\bbI + \bbD_\mathrm{visc}\Dv+\Fe\Lp\big){:}\Dv }{$=:p_\mathrm{mech}$ = the mechanical power (in W/m$^3$)}\!\!\!\!\!
-\DIV\Big(\!\!\!\lineunder{\!\bbK_\mathrm{heat} \nabla\frac1\Theta}{$\ \ \ =:\bfj_\mathrm{ener}$}\!\!\!\Big)\,.
\end{align}
Here we used that, because of the choice $w=e$ and $E'_{\Fe}=0$, the Cauchy stress 
tensor takes the form $\bfSigma_\mathrm{Cauchy}=-\Theta S'_{\Fe} \Fe^\top-\Theta S\bbI$,
see \eqref{eq:GEN.CauchyStress}. This establishes \eqref{eq:LocalEnergy} and shows that
the internal variables $\alpha$ and $\beta$ do not contribute to the energy flux,
because the assumed block structure of $\calR^*_\mathrm{simple}$ did not allow for cross
diffusion.\medskip

Finding the local balance law for the entropy density $S$ is most easily done
by making 
\[
\text{the specific choice} \quad w=s, \qquad \text{i.e.} \quad
S(\Fe,\alpha,\beta,s)=s.
\]
Then, we find $S'_{\Fe}=\bm0$, $S'_\alpha=0=S'_\beta$, $S'_s=1$ and
$E'_s=\Theta$. We also recall that $V^S_\mathrm{Ham}$ is such that for the
purely Hamiltonian 
system the entropy density is simply transported as a extensive variable 
along the vector field $\vv$, see \eqref{eq:S.transported}. Thus, the 
term $V^w_\mathrm{irr}$ will produce the terms for entropy production
$\sigma_\mathrm{prod}$ and for the entropy flux $\bfj_\mathrm{entr}$. 

Assembling the terms from the derivation in Sections \ref{su:GEN.Hamil} and
\ref{su:GEN.diss}, we find
\begin{align}\nonumber
\plt s + \DIV(s\vv)
&= \frac1\Theta\Big( \Dv{:}\bbD_\mathrm{visc}
  \Dv {+} E'_{\Fe}{:}(\Fe\Lp) {-} E'_\alpha V^\alpha_\mathrm{irr} 
  {-} E'_\beta V^\beta_\mathrm{irr} {-}
  \DIV\big(\bbK_\mathrm{heat}\nabla\frac1\Theta\big)\Big) 
\\\nonumber
&= \sigma_\mathrm{prod} - \DIV\big( \bfj_\mathrm{entr} \big) \quad \text{with} 
\\\nonumber\sigma_\mathrm{prod}
&= \frac1\Theta \Dv{:}\bbD_\mathrm{visc}  \Dv+
\bbM\bfSigma_\mathrm{Mandel}{:}\pl_{\bfeta^{\Fe}}
R^*_\mathrm{plast}\big(\bbM\bfSigma_\mathrm{Mandel} \big)
\\&\nonumber\  +
\nabla\frac{\mu^\alpha}\Theta{\cdot} \bbA_\mathrm{diff}
\nabla\frac{\mu^\alpha}\Theta+A_\mathrm{source}\Big(\frac{\mu^\alpha}\Theta\Big)^2\!
+\nabla\frac{\mu^\beta}\Theta{\cdot}\bbB_\mathrm{diff}\nabla\frac{\mu^\beta}\Theta
 +  \nabla\frac1\Theta{\cdot} \bbK_\mathrm{heat}\nabla\frac1\Theta\,,  
\\
\bfj_\mathrm{entr}&=\frac{\mu^\alpha}\Theta\bbA_\mathrm{diff}
\nabla\frac{\mu^\alpha}\Theta+
\frac{\mu^\beta}\Theta\bbB_\mathrm{diff}\nabla\frac{\mu^\beta}\Theta
 +  \frac1\Theta\bbK_\mathrm{heat}\nabla\frac1\Theta\,,
\label{entropy-eq}\end{align}
where the chemical potentials take the form $\mu^\alpha= E'_\alpha -\Theta\!\; 0$
and $\mu^\beta= E'_\beta -\Theta\;\!0$ 
because of the choice $w=s=S$. This establishes \eqref{eq:LocalEntropy}.

\def\eta{s}

\section{How the system \eq{the-system} arises from {\smaller GENERIC}}\label{sec-deriv}

A basic ingredient of the model \eq{the-system}
is that the free energy is taken referential, whereas the actual free energy 
$\psi = E{-}\Theta S$ is better fitted to 
directly fitted to the {\smaller GENERIC} approach in Sect.~\ref{sec-GENERIC},
but perhaps less standard in engineering or physics, see \eq{psi-vs-psi} for the
relation. As in Section~\ref{sec-GENERIC}, we use the {\it actual entropy}
$\eta=-\psi_\theta'(\Fe,\ZETA,\theta)$, the {\it actual chemical potential}
$\mu=\psi_\ZETA'(\Fe,\ZETA,\theta)$, and the {\it actual internal energy}
(Gibbs' relation) $e=\psi+\theta\eta$.

We willstart from the system \eq{system}--\eq{system-data} with the choice
$w=e$ to see the energy balance in its standard form. However, we will then use
the relation $e=E(\Fe,z,\theta)$ for formulating the equations in terms of the
temperature. Moreover, we will simplify the approach in Sect.~\ref{sec-GENERIC}
by assuming further
\begin{subequations}
\begin{align}
&{\rm tr}\Lp=0\,,\ \ \ \ 
\bbM={\rm dev}:\R^{d\times d}\to\R_{\rm dev}^{d\times d}:
A\mapsto A-\frac{{\rm tr}A}d\bbI\,,\ \ 
\text{ and }\ \ A_\mathrm{source}=0\,.
\intertext{Taking $R_{\rm plast}$ whose convex conjugate $R_{\rm plast}^*$ occurs in 
\eq{system-data-b}, we consider the dissipation potential $r$ used in
\eq{the-system} as}
&r(\ZETA,\theta;\Dv,\Lp)=\frac12\Dv{:}\bbD_\mathrm{visc}\Dv
+R_{\rm plast}\Big(\ZETA,\theta;\frac{\Lp}\theta\Big)
\ \ \text{ with }\ \bbD_\mathrm{visc}=\bbD_\mathrm{visc}(\ZETA,\theta)\,.
\label{data-r}
\end{align}\end{subequations}
It is to note that \eq{data-r} turns the kinetic equation (flow rule) \eq{system-data-b}
into $[R_{\rm plast}]_{\Lp}'\!(\ZETA,\theta,\Lp/\theta)={\rm dev}\bfSigma_\mathrm{Mandel}$
with the Mandel stress
$\bfSigma_\mathrm{Mandel}=\Fe^\top\psi_{\Fe}'(\Fe,\ZETA,\theta)$,
where we used that $(\bbM^*)^{-1}$ is the identity on $\R_{\rm dev}^{d\times d}$
where $\Lp$ is valued. Again using \eq{data-r}, we can write it as
$r_{\Lp}'\!(\ZETA,\theta;\Dv,\Lp)={\rm dev}\bfSigma_\mathrm{Mandel}$.
When using \eq{system}--\eq{system-data}, we realize that $E-\Theta S=\psi$.
This altogether turns \eq{system}--\eq{system-data} into the system:\def\W{e}
\begin{subequations}\label{the-system+}
\begin{align}
&\nonumber  \pdt{}(\varrho \vv)+\DIV(\varrho\vv{\otimes}\vv)
=\varrho\GRAVITY+\DIV\big(\bfSigma_\mathrm{Cauchy}
{+}\bfSigma_\mathrm{dissip}\big)\ \ \text{ with }\ \bfSigma_\mathrm{dissip}=
r_\Dv'(\ZETA,\theta;\Dv,\Lp)
\\&\hspace{7em}\text{and }\ 
\bfSigma_\mathrm{Cauchy}=\psi_{\Fe}'(\Fe,\ZETA,\theta)\Fe^\top\!
+\big(\psi(\Fe,\ZETA,\theta){-}\,\SW\ZETA\psi_\ZETA'(\Fe,\ZETA,\theta)\big)\bbI\,,
\label{momentum-eq+}
\\&\DT\Fe=(\nabla\vv)\Fe-\Fe\Lp\,,
\label{evol-of-E++}
\\&
\DT\ZETA+\SW\ZETA\,\DIV\,\vv=
\DIV\Big(\bbK_\mathrm{diff}\nabla\dfrac\mu\theta\Big)
\label{system-diffusion-eq+}
\ \ \ \text{ with }\ \mu=\psi_\ZETA'(\Fe,\ZETA,\theta)\,,
\\&r_{\Lp}'\!(\ZETA,\theta;\Dv,\Lp)={\rm dev}\,\bfSigma_\mathrm{Mandel}
\ \ \ \text{ with }\ \bfSigma_\mathrm{Mandel}=\Fe^\top\psi_{\Fe}'(\Fe,\ZETA,\theta)\,,
\label{inelastic-flow+}
\\&\nonumber\hspace*{-0em}
\pdt\W+\DIV\big(\W\vv\big)=p_{\rm heat}-\DIV\Big(\bbK_\mathrm{heat}\nabla\frac1\theta\Big)
+\big(\psi_{\Fe}'(\Fe,\ZETA,\theta)\Fe^\top\!{+}\psi(\Fe,\ZETA,\theta)\bbI\big)
{:}\nabla\vv
\\&\nonumber\hspace{12em}
-\Fe^\top\psi_{\Fe}'(\Fe,\ZETA,\theta){:}\Lp
-\SW\ZETA\psi_\ZETA'(\Fe,\ZETA,\theta)\,\DIV\,\vv
\\&\nonumber\hspace*{5em}
\text{ with the heat production rate }\
p_\mathrm{heat}=\bfSigma_\mathrm{dissip}{:}\Dv+r_{\Lp}'\!(\ZETA,\theta;\Dv,\Lp){:}\Lp
\\[-.2em]&\hspace{5em}
\text{ and the internal energy }\
\W=\psi(\Fe,\ZETA,\theta)-\theta\psi_\theta'(\Fe,\ZETA,\theta)\,,
\label{Euler-large-heat-eq+}
\end{align}
\end{subequations}
with $\SW$ from \eq{switch}; note that we wrote \eq{system-c} and
\eq{system-d} in a unified way so that $\ZETA$ can be both $\alpha$
or $\beta$ and correspondingly $\bbK_\mathrm{diff}$ can be
$\bbA_\mathrm{diff}$ or $\bbB_\mathrm{diff}$.

To inspect the internal energy equation \eq{energy-eq} and
the entropy equation \eq{entropy-eq}, let us realize that
\begin{subequations}\label{data}\begin{align}
&p_\mathrm{mech}=\big( \bfSigma_\mathrm{Cauchy}-\ZETA(E_\ZETA'{-}\theta S_\ZETA')\bbI
+\bfSigma_\mathrm{dissip}\big){:}\Dv\,,
\\&\bfj_\mathrm{ener}=\bbK_\mathrm{heat}\nabla\frac1\theta\,,
\\&\nonumber
\sigma_\mathrm{prod}=\frac{p_{\rm heat}}\theta
+\nabla\frac\mu\theta{\cdot}\bbK_\mathrm{diff}
\nabla\frac\mu\theta+\nabla\frac1\theta{\cdot}\bbK_\mathrm{heat}\nabla\frac1\theta
\ \ \\[-.2em]&\hspace{15em}\text{with }p_{\rm heat}=
r_{(\Dv,\Lp)}'(\ZETA,\theta;\Dv,\Lp){:}(\Dv,\Lp)\,,  
\label{data-pheat}\\[-.5em]&
\bfj_\mathrm{entr}=\frac\mu\theta\bbK_\mathrm{diff}
\nabla\frac\mu\theta+\frac1\theta\bbK_\mathrm{heat}\nabla\frac1\theta\,.
\end{align}\end{subequations}
As articulated in Section~\ref{sec-GENERIC}, these general equations lead
to the total energy balance and the total
enetropy balance, respectively. For this, we need to consider a fixed
domain $\varOmega\subset\R^d$ and to prescribe some boundary conditions;
for simplicity, we use impermeable boundary by prescibing
\begin{subequations}\label{BC}\begin{align}\label{BC1}
&\vv{\cdot}\nn=0\,,\ \ \ \ \jj_{\rm ener}{\cdot}\nn=0\,,\ \ \ \text{ and }\ \ \
\jj_{\rm entr}{\cdot}\nn=0\ \ \text{ on }\ \pl\varOmega\,
\intertext{with $\nn$ denoting the outward normal to the boundary
$\pl\varOmega$ of the domain $\varOmega$. Besides, we prescibe}
\label{BC2}
&(\bfSigma_\mathrm{Cauchy}{+}\bfSigma_\mathrm{dissip})\nn=\bm0 \,,\ \ \ \
\nabla\theta{\cdot}\nn=0\,,\ \ \text{ and }\ \ \
\nabla\mu{\cdot}\nn=0\ \ \text{ on }\ \pl\varOmega\,.
\end{align}\end{subequations}
Actually, the last two conditions in \eq{BC2} allow the last condition \eq{BC1} to
be omitted.

Integrating \eq{energy-eq} over $\varOmega$ and using the specific
form of $p_{\rm mech}$ and the momentum equation \eq{momentum-eq}
tested by $\vv$ together with the mass continuity equation \eq{cont-eq}
above, we obtain {\it total energy balance}:
\begin{align}\nonumber
\!\frac{\d}{\d t}\int_\varOmega\! e\,\d\xx&=
\int_\varOmega p_\mathrm{mech}\,\d\xx\stackrel{\eq{energy-eq}}{=}\!\!\int_\varOmega(\bfSigma_\mathrm{Cauchy}{+}\bfSigma_\mathrm{dissip}){:}\Dv\,\d\xx
\\&=\int_\varOmega\Big(\varrho\GRAVITY
-\pdt{}(\varrho \vv)-\DIV(\varrho\vv{\otimes}\vv)\Big){\cdot}\vv\,\d\xx
\stackrel{\eq{cont-eq}}{=}\!\!\int_\varOmega\varrho\GRAVITY{\cdot}\vv\,\d\xx
-\frac{\d}{\d t}\int_\varOmega\frac\varrho2|\vv|^2\,\d\xx\,.
\label{tot-engr}\end{align}
It shows the conservation of the total energy
$\int_\varOmega \frac12\varrho|\vv|^2+e\,\d\xx$ if the gravitation would be
neglected, i.e.\ if $\GRAVITY=0$.

Integrating \eq{entropy-eq} over $\varOmega$ and using \eq{data-pheat},
we obtain {\it total entropy balance}:
\begin{align}
\frac{\d}{\d t}\int_\varOmega \eta\,\d\xx=\int_\varOmega\sigma_{\rm prod}\,\d\xx
&=\int_\varOmega\frac{p_{\rm heat}}\theta
+\nabla\frac\mu\theta{\cdot}\bbK_\mathrm{diff}\nabla\frac\mu\theta
+\nabla\frac1\theta{\cdot} \bbK_\mathrm{heat}\nabla\frac1\theta
\,\d\xx\,.
\label{ent-balance}\end{align}

Taking \eq{entropy-eq} multiplied by $\theta$, i.e.\
$\theta\DT\eta=\theta\sigma_{\rm prod}-\theta\DIV\,\jj_{\rm entr}
-\theta\eta\DIV\,\vv$, and substituting $\eta=-\psi_\theta'(\Fe,\ZETA,\theta)$,
we can see the heat equation in the ``engineering form'' in terms of
temperature $\theta$ as an intensive variable:
\begin{align}\nonumber
c(\Fe,&\ZETA,\theta)\DT\theta
=\theta\sigma_{\rm prod}-\theta\DIV\,\jj_{\rm entr}
+\theta\psi_{\Fe\theta}''(\Fe,\ZETA,\theta){:}\DT\Fe
+\theta\psi_{\ZETA\theta}''(\Fe,\ZETA,\theta)\DT\ZETA
+\theta\psi_\theta'(\Fe,\ZETA,\theta)\,\DIV\,\vv
\\\nonumber
&\!\!\!\stackrel{\scriptsize\rm(\ref{data}c,d)}{=}p_{\rm heat}
+\theta\nabla\frac1\theta{\cdot} \bbK_\mathrm{heat}\nabla\frac1\theta
+\theta\nabla\frac\mu\theta{\cdot} \bbK_\mathrm{diff}\nabla\frac\mu\theta
-\theta\DIV\Big(\frac\mu\theta \bbK_\mathrm{diff}\nabla\frac\mu\theta+\frac1\theta\bbK_\mathrm{heat}\nabla\frac1\theta\Big)
\\\nonumber&\qquad\qquad
+\theta\psi_{\Fe\theta}''(\Fe,\ZETA,\theta){:}\DT\Fe
+\theta\psi_{\ZETA\theta}''(\Fe,\ZETA,\theta)\DT\ZETA
+\theta\psi_\theta'(\Fe,\ZETA,\theta)\,\DIV\,\vv
\\\nonumber
&\!\stackrel{\scriptsize\eq{system-diffusion-eq}}{=}p_{\rm heat}
-\DIV\Big(\bbK_\mathrm{heat}\nabla\frac1\theta\Big)
+\theta\psi_{\Fe\theta}''(\Fe,\ZETA,\theta){:}\DT\Fe
\\\nonumber&\hspace{8em}
+\big(\theta\psi_{\ZETA\theta}''(\Fe,\ZETA,\theta){-}\mu\big)\DT\ZETA
+\big(\theta\psi_\theta'(\Fe,\ZETA,\theta){-}\SW\ZETA\mu\big)\,\DIV\,\vv
\\\nonumber
&\!\stackrel{\scriptsize\eq{evol-of-E}}{=}p_{\rm heat}
-\DIV\Big(\bbK_\mathrm{heat}\nabla\frac1\theta\Big)+
\theta\big(\psi_{\Fe}'(\Fe,\ZETA,\theta)\Fe^\top{+}\psi(\Fe,\ZETA,\theta)\bbI\big)_\theta'{:}\nabla\vv
\\\nonumber
&\hspace{1em}-\theta\Fe^\top\psi_{\Fe\theta}''(\Fe,\ZETA,\theta){:}\Lp
-\big(\psi(\Fe,\ZETA,\theta)-\theta\psi_\theta'(\Fe,\ZETA,\theta)\big)_\ZETA'\DT\ZETA-\SW\ZETA\psi_\ZETA'(\Fe,\ZETA,\theta)\,\DIV\,\vv
\\[.1em]&\hspace{10em}\text{with the heat capacity }\
c(\Fe,\ZETA,\theta)=-\theta\psi_{\theta\theta}''(\Fe,\ZETA,\theta)\,.
\label{Euler-heat-eq}
\end{align}

Then we use the kinematic equation \eq{evol-of-E} with the algebra
$A{:}(BC)=(B^\top A){:}C=(AC^\top){:}B$
so that, using also $\W=\psi-\theta\psi_\theta'$ and realizing that
$\W_\theta'(\Fe,\ZETA,\theta)=\psi_\theta'-(\theta\psi_\theta')_\theta'=
-\theta\psi_{\theta\theta}''$ is the heat capacity $c$, we can evaluate the left-hand
side of \eq{energy-eq} as
\def\W{e}
\begin{align}\nonumber
\pdt\W&+\DIV(\W\vv)=\DT\W+\W\DIV\,\vv
\\[-.3em]&\nonumber=\W_{\Fe}'(\Fe,\ZETA,\theta){:}\DT\Fe
+\W_{\ZETA}'(\Fe,\ZETA,\theta)\DT\ZETA+\W_\theta'(\Fe,\ZETA,\theta)\DT\theta
+\W(\Fe,\ZETA,\theta)\DIV\,\vv
\\\nonumber&=
c(\Fe,\ZETA,\theta)\DT\theta+
\big(\psi_{\Fe}'(\Fe,\ZETA,\theta){-}\theta\psi_{\Fe\theta}''(\Fe,\ZETA,\theta)\big){:}
\DT\Fe
\\\nonumber&\hspace{6.em}
+\big(\psi_{\ZETA}'(\Fe,\ZETA,\theta){-}\theta\psi_{\ZETA\theta}''(\Fe,\ZETA,\theta)\big)\DT\ZETA+
\big(\psi(\Fe,\ZETA,\theta){-}\theta\psi_{\theta}'(\Fe,\ZETA,\theta)\big)
\DIV\,\vv\,,
\\\nonumber
&\!\stackrel{\scriptsize\eq{evol-of-E}}{=}
c(\Fe,\ZETA,\theta)\DT\theta+
\big(\psi_{\Fe}'(\Fe,\ZETA,\theta)\Fe^\top\!{+}\psi(\Fe,\ZETA,\theta)\bbI\big){:}\nabla\vv
-\theta\big(\psi_{\Fe}'(\Fe,\ZETA,\theta)\Fe^\top\!{+}\psi(\Fe,\ZETA,\theta)\bbI\big)_\theta'{:}\nabla\vv
\nonumber\\&\hspace{6.em}
-\Fe^\top\big(\psi_{\Fe}'(\Fe,\ZETA,\theta){-}\theta\psi_{\Fe\theta}''(\Fe,\ZETA,\theta)\big){:}\Lp+\big(\psi_{\ZETA}'(\Fe,\ZETA,\theta){-}\theta\psi_{\ZETA\theta}''(\Fe,\ZETA,\theta)\big)\DT\ZETA\,.
\nonumber\end{align}
Thus, substituting for $c(\Fe,\ZETA,\theta)\DT\theta$ from
\eq{Euler-heat-eq}, we obtain just \eq{Euler-large-heat-eq+}.

In terms of the referential free energy $\uppsi$ instead of the actual free
energy $\psi$ used for {\smaller GENERIC}, the system (\ref{the-system+}a-d)
with \eq{Euler-large-heat-eq+} written in the form \eq{Euler-heat-eq} and
completed with the mass continuity equation transforms into the original system
\eq{the-system}.  Indeed, recalling \eq{psi-vs-psi}, it is to be noted that the
Cauchy stress rewrites as
\begin{align}\nonumber
\bfSigma_\mathrm{Cauchy}&=\psi_{\Fe}'(\Fe,\ZETA,\theta)\Fe^\top+\big(\psi(\Fe,\ZETA,\theta){-}\SW\ZETA\psi_\ZETA'(\Fe,\ZETA,\theta)\big)\bbI
\\&=\dfrac{\uppsi_{\Fe}'(\Fe,\ZETA,\theta)\Fe^\top\!-\SW\ZETA\uppsi_\ZETA'(\Fe,\ZETA,\theta)\bbI}{\det\Fe}\,,
\end{align}
which reveals the conservative part of the Cauchy stress in \eq{momentum-eq}.
Here we used the algebra $F^{-1}={\rm Cof}^\top\!F/\!\det F$ and
the calculus $\det'(F)={\rm Cof}\,F$ for
\begin{align}\nonumber
\frac{\uppsi_{\Fe}'(\Fe,\ZETA,\theta)}{\det\Fe}&\Fe^\top
=\frac{\uppsi_{\Fe}'(\Fe,\ZETA,\theta)
-\uppsi(\Fe,\ZETA,\theta)\Fe^{-\top}\!\!\!\!}{\det\Fe}\Fe^\top\!+
\frac{\uppsi(\Fe,\ZETA,\theta)}{\det\Fe}\bbI
\\&\nonumber\hspace*{-2em}
=\bigg(\frac{\uppsi_{\Fe}'(\Fe,\ZETA,\theta)}{\det\Fe}
-\frac{\uppsi(\Fe,\ZETA,\theta){\rm Cof}\Fe}{(\det\Fe)^2}\bigg)\Fe^\top\!+
\frac{\uppsi(\Fe,\ZETA,\theta)}{\det\Fe}\bbI
\\& \hspace*{-2em}
=\Big[\frac{\uppsi(\Fe,\ZETA,\theta)}{\det\Fe}\Big]_{\Fe}'\Fe^\top\!
+\frac{\uppsi(\Fe,\ZETA,\theta)}{\det\Fe}\bbI
 \ \overset{\text{\eq{psi-vs-psi}}}{=} \ 
\psi_{\Fe}'(\Fe,\ZETA,\theta)\Fe^\top\!+\psi(\Fe,\ZETA,\theta)\bbI\,.
\label{referential-conservatives}\end{align}
This calculation is to be used also for \eq{Euler-heat-eq}, which
then gives the temperature equation \eq{Euler-large-heat-eq}.
The diffusion equation \eq{system-diffusion-eq} arises
directly from \eq{system-diffusion-eq+} when considering
\eq{psi-vs-psi}, while \eq{inelastic-flow+} relies on ${\rm tr}\Lp=0$.

\begin{remark}[Energy in J/kg.]\label{rem-Einstein}\upshape
  Another variant of the referential free energy $\uppsi$ often used in
  literature is to express it in terms of J/kg instead of J/m$^3$, i.e.\ \
  Joule per a referential volume. This is actually the original Einstein idea
  of expressing elastic moduli in such physical units, i.e.\ in fact in
  (m/s)$^2$, which is compatible with his famous formula ``e/m=c$^2$''. Then
  the actual free energy is, instead of \eq{psi-vs-psi}, given by
  $\psi(\Fe,\ZETA,\theta)=\varrho\uppsi(\Fe,\ZETA,\theta)$.  The corresponding
  variant of the system \eq{the-system} arises by replacing the factor
  $1/\det\Fe$ by $\varrho$ everywhere in (\ref{the-system}b,d--f).
\end{remark}

\section{Examples of application}\label{sec-appl}

The general model \eq{the-system} has many applications in finitely-strained continuum
thermo-mechanics. Let us illustrate it on two examples only, both involving 
interesting phenomena studied in continuum mechanics, namely various
{\it phase transitions}. They can be primarily {\it volumetric} or {\it deviatoric}
(isochoric), the former one being induced by high pressure \cite{BhaSin12PISP}.

Another highly developed part of continuum mechanics is {\it poromechanics}
in a broader sense, cf.\ \cite{Cous04P,Cush97PFHP,Boer05TCMP}.
Effects as advection and diffusion are typically accompanied or coupled with
mechanical swelling or squeezing, i.e.\ volumetric changes. In this way, the
coupling with the mentioned volumetric phase transitions can occur.
Substantial volumetric changes are especially during various solid-solid phase
transitions with applications in geo-engineering (e.g.\ thawing of permafrost),
petrology, materials science (hydrogels or metal plasticity and hydrogenation
metal-hydride transition during hydrogen diffusion in metals),
and geophysics (mantle dynamics and dewatering/dehydration of rocks).

We will briefly illustrate this rich variety of applications
with two examples, one with the volumetric phase transitions
and the other with the deviatoric phase transitions.
There is a hierarchy of various kinds of models. Here we confine
ourselves to a rather phenomenological Biot-type model.

\subsection{Earth's mantle dynamics, phase transitions and (de)hydration}

The first example concerns with the volumetric phase transitions in
Earth's mantle associated with mass density jumps at specific depths,
combined also with ``water'' diffusion. In contrast to the general model
devised in \cite{Roub21TCMP} which uses additive decomposition
and a linearized convective model as most frequently used in geophysics,
here we used a fully nonlinear model with a multiplicative decomposition
and a thermodynamically consistently formulated diffusion, as in
the system \eq{the-system}.  

In fact, the water in the mantle is chemically bonded rather than real liquid water
in pores. Actually, there are many hydrous minerals \cite{Ohta21HDEI}.
Altogether, most of the water on the planet Earth is not in the oceans but in its
mantle, the rough estimation being mostly from 1 to 10 times the ocean mass. 
A particularly large amount of water (about 1--2\,wt\%) is located in the
mantle transition zone between 410\,km and 660\,km below Earth's surface.
On these interfaces (and also some others), phase transitions are volumetric.
We confine ourselves on unhysteretic situations governed by a model
with a convex energy.

A prominent modelling activity in geophysical mantle dynamics is the
applications of descending slabs undergoing dehydration.  More in detail, it
originates when oceanic plates subduct continental plates.  Being relatively
cold and dense, they are descending deep into the Earth's mantle. They are
originally hydrated by water from oceans, with water content being typically
more than 1\%.  On their route into Earth's interior, they are subjected by
ever increasing pressure and undergo a series of phase transitions.  The most
important and sharp transitions are under pressures about 14\,GPa and 24\,GPa
which are at the mentioned depth about 410\,km and 660\,km below
Earth's surface, where transitions from olivine to wadsleyite and ringwoodite
to bridgmanite/magnesiowustite are undergoing, respectively. Both these phase
transitions are accompanied by quite sharp increase of mass density.

Descending slabs can either stagnate on the 660\,km-discontinuity or continue
their journey into the lower mantle up to the core-mantle boundary
deep in 2.900\,km. While undergoing their phase transition reaching 24\,GPa
pressure at 660\,km depth, they tend to dehydrate while hydrating the
surrounding mantle, cf.\ \cite{Hirs06WMDE,Ohta21HDEI}. In this way, the mantle
transition zone 410--660\,km is hydrated so that it now represents the
by-far largest reservoir of (chemically bound) water on the planet Earth.

Understanding deep slab processes in the mantle and modelling fluid
transport in subduction zones was articulated as one of major 
directions in geophysical modelling \cite{GeryFDSM}. For
pure slab descent (without dehydration), incompressibility is usually adopted while
taking into account temperature effects as in
\cite{BehCiz08LWCS,Bill08MDSS,CiHuBe07SDSS,RiMoRa07SDFM,TSGS93EEPT}
or the dehydration and phase transitions are incorporated very phenomenologically
or rather indirectly \cite{RiMoRa07SDFM,RMHC04SSZW,SJYF18STRS}, which is the
main difference from our isothermal approach. In particular, ``slab dehydration
also seems to be a possible mechanism to switch from stagnation to penetration''
\cite{ArGoHu17SSTZ} in addition to the viscosity increase in regions where seismic
tomography has imaged slab stagnation \cite{RuLeLB15VJEM}.

The conservative part of the model can use a neo-Hookean Biot-type free energy
\begin{align}
\uppsi(\Fe,\ZETA,\theta)=\KAPA(J,\theta)+G\big(J^{-2/d}{\rm tr}(\Fe\Fe^\top){-}d\big)
  +\theta\frac B2|\ZETA{+}b(J{-}J_\text{\sc t}^{})|^2
+c\theta(1{-}{\rm ln}\theta)+\delta_{[0,1]}^{}(\ZETA)\,,
\label{ansatz}\end{align}
where $J=\det\Fe$ and $\delta_{[0,1]}^{}$ denotes the indicator function
of the interval $[0,1]$; therefore $\ZETA$ is understood as an intensive
variable valued in $[0,1]$. Note that \eq{ansatz} is consistent with the ansatz
\eq{one-ansatz}. Another important ingredient is a proper viscosity. Although
there are quantitatively not much precise ideas about the radial viscosity profile
in Earth mantle, there seems a general agreement that there is a substantial
increase of viscosity at the 660km-interface, cf.\ \cite{KauLam00MDPR}
for comparison of various profiles in literature.
The Maxwellian-type viscosity potential $\nu_2(\ZETA,\theta;\cdot)$
in \eq{dissip-pot} can be quadratic in terms of $\Lp$ (then it
models so-called {\it diffusion creep} in the mantle) or polynominal with the
exponent $p\ne2$. For $p\sim 1.3$, it is used to model so-called {\it dislocation
creep}. Such power-law fluids with $p<2$ models are known in mathematical 
fluid theory under the name of {\it shear-thickening non-Newtonian} viscous,
or here viscoelastic (in the Jeffreys rheology) {\it fluids}. Such
shear-thickening rheology is an important modelling aspect leading to 
certain ``lubrication'' effects that facilitate an easy descent of cold slabs
in the surrounding warmer Earth mantle, cf.\ \cite{Bill08MDSS,CBSM12VELM,SSSY77VDPE}.

Physically, there are various simplifications when having in mind the
geophysical applications mentioned above.  In particular, we do not distinguish
between bound water and free water, assuming that it can be roughly modelled by
varying water mobility suitably, while for a ``multi-water'' variant we refer to
\cite{ZHPV23PMMR}. Anyhow, our thermo-hydro-mechanical model we have devised
seems well competitive with the usual geophysical models for rock dehydration
in subducting slabs like \cite{NaIwNa16EWTS,ZaPeTh23GFRF,ZHPV23PMMR} which
typically involve many simplifications as the mentioned incompressibility or
ignoring temperature variations etc.

\begin{figure}[ht]
\begin{center}
\psfrag{pT}{\scriptsize $p_\text{\sc t}^{}$}
\psfrag{p}{\scriptsize $p$ [GPa]}
\psfrag{det F}{\scriptsize $\det\FF$}
\psfrag{f}{\scriptsize $\Kapa$}
\psfrag{r}{\scriptsize $\varrho=\varrho_\text{\sc r}^{}/[\widehat\Kapa']^{-1}(-p)$}
\psfrag{f,}{\scriptsize $p=-\Kapa'(\det\FF)$}
\psfrag{14GPa}{\scriptsize 14\,GPa}
\psfrag{24GPa}{\scriptsize 24\,GPa}
\psfrag{14}{\scriptsize 14}
\psfrag{24}{\scriptsize 24}
\psfrag{1}{\scriptsize 1}
\psfrag{0}{\scriptsize 0}
\psfrag{rR}{\scriptsize $\varrho_\text{\sc r}^{}$}
\psfrag{straight segment}{\scriptsize\begin{minipage}[t]{11em}straight\\[-.3em]\hspace*{-.0em}segment\end{minipage}}
\psfrag{jump2}{\scriptsize\begin{minipage}[t]{11em}jump of mass\\[-.2em]\hspace*{0em}density ($\sim5\%$)\end{minipage}}
\psfrag{jump1}{\scriptsize\begin{minipage}[t]{11em}jump of mass\\[-.2em]\hspace*{0em}density ($\sim3\%$)\end{minipage}}
\psfrag{transition}{\scriptsize\begin{minipage}[t]{11em}transition\\[-.2em]\hspace*{1em}zone\\[-.2em]\hspace*{-.6em}(410-660\,km)\end{minipage}}
\psfrag{lower}{\scriptsize\begin{minipage}[t]{11em}\hspace*{1em}lower\\[-.2em]\hspace*{1em}mantle\\[-.2em]\hspace*{-.7em}(660-2900\,km)\end{minipage}}
\psfrag{upper}{\scriptsize\begin{minipage}[t]{11em}uppermost\\[-.2em]\hspace*{-.2em}mantle and\\[-.2em]\hspace*{-.4em}astenosphere\end{minipage}}
\psfrag{Clapeyron1}{\scriptsize\begin{minipage}[t]{11em}Clapeyron\\[-.2em]\hspace*{.1em}slope 1.6\,MPa/K\end{minipage}}
\psfrag{Clapeyron2}{\scriptsize\begin{minipage}[t]{11em}Clapeyron\\[-.2em]\hspace*{.1em}slope $-$2.5\,MPa/K\end{minipage}}
\includegraphics[width=.99\textwidth]{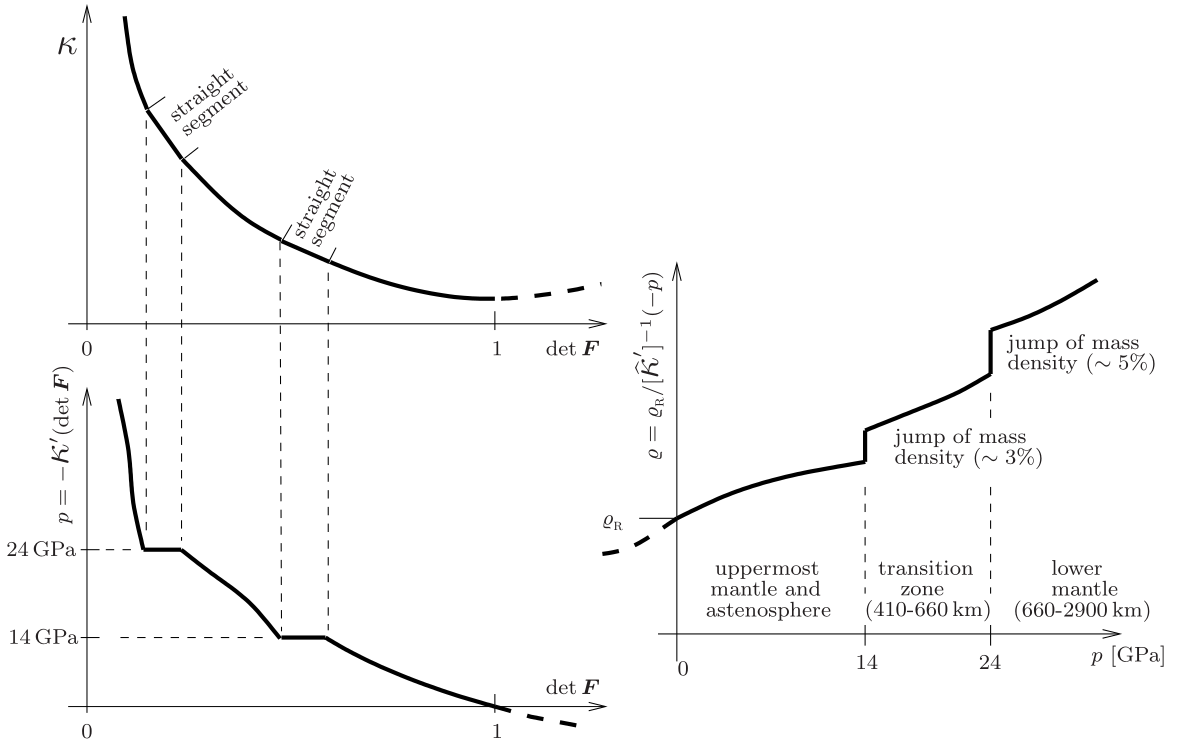}
\end{center}
\vspace*{-.6em}
\caption{
{\sl A schematic illustration of an isothermal phase transition
(rock compaction) occuring at two specific pressures 14\,GPa
(at $\sim\,$410\,km depth) and 24\,GPa (at $\sim\,$660\,km depth) in
Earth's mantle modelled by a convex function $\KAPA(\cdot,\theta)$.}}
\label{PT-rocks}
\end{figure}

\begin{remark}[State equation and phase transitions.] \upshape
Considering $B=0$ in \eq{ansatz} for a moment, the actual bulk stored energy
$\kapa(J,\theta)=\KAPA(J,\theta)/J$ with $\KAPA$ from \eq{ansatz} yields the
(actual) pressure $p=-\kapa_J'(J,\theta)$. This determines the state equation
$p=\pi(\varrho,\theta):=-\kapa_J'(\varrho_\text{\sc r}^{}/\varrho,\theta)$.
Then $\varrho$ is a function of pressure and temperature $\varrho=\rho(p,\theta)$,
namely $\varrho=\varrho_\text{\sc r}^{}/J
=\varrho_\text{\sc r}^{}/[\kapa_J'(\cdot,\theta)]^{-1}(-p)
=\varrho_\text{\sc r}^{}/[\kapa^*(\cdot,\theta)\KAPA]_p'(-p)=:\rho(p,\theta)$
where $\kapa^*(\cdot,\theta)$ denotes the convex conjugate to
$\kapa(\cdot,\theta)$ acting on $-p$. The men\-tioned mass-density discontinuities
and the corresponding volumetric phase transitions can be modelled by a convex
smooth function $\kapa(\cdot,\theta)$ which has two linear segments. 
Thus the dependence of the pressure $p=-\kapa_J'(\cdot,\theta)$ on $J$
has two horizontal plateaus (here taking the values 14\,GPa and 24\,GPa);
cf.\ Figure~\ref{PT-rocks}. Then $\rho(\cdot,\theta)$ is discontinuous with two
jumps at 14\,GPa and 24\,GPa. We should then speak rather about a set-valued
function and write the state equation as $\varrho\in\rho(p,\theta)$.
In general, the mentioned phase transitions at specific pressures thus depend
also on temperature. The thermic character of these phase transitions is
related to the so-called {\it Clapeyron slope}
$\pl p / \pl\theta=\pi_\theta' (\varrho,\theta) =
-[\kapa]_{J\theta}''(\varrho_\text{\sc r}^{}/\varrho,\theta)$. For positive
(resp.\ negative) $\pl p/\pl\theta$, the corresponding {\it phase transition}
is {\it exothermic} (resp.\ {\it endothermic}). Adiabatic effects due to volume
changes still contribute slightly to this thermic character but, in fact, this
contribution is rather minor in the phase transitions in Earth's mantle.
Specifically, the transition from lower-density ringwoodite (above 660\,km) to
higher-density perovskite (below 660\,km) is endothermic with the Clapeyron
slope $-$2.5\,MPa/K, while the transition from olivine (above 410\,km) to
wadsleyite (below 410\,km) is opposite with the Clapeyron slope 1.6\,MPa/K. This
temperature dependence is not depicted in Figure~\ref{PT-rocks}, however. 
\end{remark}

\subsection{Martensitic phase transition with plasticity and diffusion}

Our second example is the martensitic phase transformation in so-called shape-memory
alloys, extensively studied for decades in hundreds of articles and also in
monographs such as \cite{AbeKno06EPT,Bhat03MMWF,PitZan03CMPT}.
At the single-crystal level, it is modelled by a nonconvex multi-well
free energies with $1\,{+}\,n$ wells as orbits of the types ${\rm SO}(d)\FF_{\rm i}$
with $\FF_0=\bbI$ for the cubic austenite and $\FF_i$ with $i=1,..,,n$ for
$n$ lower symmetrical variants of martensite; in the 3-dimensional case,
$n=3$ for tetragonal, $n=6$ for orthorhombic, or $n=12$ for monoclinic
martensite. The transition between particular martensitic phases (variants),
i.e.\ re-orientation of martensite, is isochoric (i.e.\ volume preserving)
and the transition between cubic austenite and a particular martensitic variant
is (almost) isochoric, too. For this, $\det\FF_i=1$ is assumed and
the potential $\psi(\cdot,\theta)$ is multi-well in its deviatoric part only.
A certain (slightly academical) example can be
\begin{align}
\psi(\Fe,\theta)=\KAPA(\det\Fe,\theta)+\min_{i=0,...,n}\bigg(
G_i\Big(\frac{{\rm tr}(\Fe\Fe^\top\!\FF_i^{-\top}\!\FF_i^{-1})}{(\det\Fe)^{2/d}}
-d\Big)-c_i\theta\,{\rm ln}\frac\theta{\theta_\text{\sc t}^{}}\bigg)
\end{align}
with the transition
temperature $\theta_\text{\sc t}^{}$ at which austenite is energetically
equilibrated with martensite and $c_i$ the heat capacities of particular
phases; typically $c_0>c_1=...=c_n$, which ensures that austenite is
energetically dominant for $\theta>\theta_\text{\sc t}^{}$ and vice versa.

This isochoric phase transition can be accompanied by another isochoric
process, {\it plasticity}. Plasticity in shape-memory alloys is a studied
phenomenon in materials science, see e.g.\ \cite{HSSS18PDAC,SSSS18CMTP}; for a
model in a Lagrangian isothermal quasistatic situation see \cite{KruZim11MSMA}.

The transitions between these $1\,{+}\,n$ phases as well as the plasticity are 
activated processes and, in order to evolve, they need (and dissipate)
some specific activation energy. As in \cite{KrMiRo05MMES}, the former one
can be modelled by a 1-homogeneous contribution to the dissipation potential
involving a nonlinear ``phase indicator'' function 
$\lambda:\R^{d\times d}\to\triangle\subset\R^{1+n}$
where $\triangle=\{(\lambda_0,...,\lambda_n);\ \lambda_i\ge0,\ \sum_{i=0}^n\lambda_i=1\}$
denotes the so-called Gibbs simplex. For $\Fe$ in the mentioned
wells, $\lambda(\Fe)$ takes values in on of the vertex of $\triangle$.
The corresponding term in the dissipation potential is then
$|\DT{\overline{\lambda(\Fe)}}|$, as devised in \cite{Roub04NHAC} in the
Lagrangian setting. Another contribution $\sigma|\Lp|$ models the plasticity
with $\sigma=\sigma(\theta)>0$ playing the role of a yield stress needed to
activate the plastification. To put this contribution into the form
$r(\Fe,\ZETA,\theta;\nabla\vv,\Lp)$ acting on the rates $\nabla\vv$ and
$\Lp$ as in \eq{dissip-pot}, we use \eq{evol-of-E} and the calculus
$|\DT{\overline{\lambda(\Fe)}}|=|\lambda'(\Fe)\DT\Fe|=|\lambda'(\Fe)((\nabla\vv)\Fe{-}\Lp\Fe)|$. Overall, the nonsmooth dissipation potential
$r(\Fe,\theta;\cdot,\cdot)$ reads as
\begin{align}\label{SMA-dissip}
r(\Fe,\theta;\LL,\Lp)=\big|\lambda'(\Fe)(\LL\Fe{-}\Lp\Fe)\big|
+\sigma(\theta)|\Lp|\ \ \ \text{ with $\LL$ a placeholder for $\nabla\vv$}
\,.
\end{align}
Of course, the partial derivative occurring in \eq{momentum-eq}
and \eq{inelastic-flow+} should then be the convex subdifferentials.
Yet, the potential depending on $\Dv$ is now to be generalized for
the full velocity gradient $\LL=\nabla\vv$ and, moreover,
the general coupling between $\LL$ and $\Lp$ in \eq{SMA-dissip}
does not comply with the (slightly simplified) analysis in
Sect.\,\ref{su:GEN.Hamil} which then should be enhanced by allowing more
cross-effects.

\begin{remark}[Metal-hydride phase transition.]\upshape
The deviatoric martensitic phase tran\-sition and plasticity can even be
combined with the diffusion, particularly of hydrogen, and the associated
volumetric transition, known as the {\it metal-hydride phase transition}.
For experimental evidence and small-strain models we refer
e.g.\ to \cite{Jian22EHSE,SSWA89HSDS} and
\cite{LBGB16MHEA,LeHaGa17CMHD,LHGS20CDMM,UBLB19MHET}, respectively.
\end{remark}

\section{Notes to analysis.}\label{sect-notes}

There is a certain agreement that models of solids at large strains analytically
require usage of some higher-gradient theories, unless
some very week (e.g.\ of measure-valued-type) solution concepts are used.
It is used in the conservative part for the Lagrangian setting while rather in
the dissipative part for the Eulerian setting. The latter case, which
can be used here, is referred as {\it multipolar continua},
devised by Mindlin \cite{Mind64MSLE} and Toupin \cite{Toup62EMCS}
and later used e.g.\ by \cite{FriGur06TBBC,Neca94TMF,NeNoSi89GSIC,NecRuz92GSIV}.
Such multipolar modification (regularization), which expands the dissipative
stress by a term like $-\DIV(\nu|\nabla^2\vv|^{p-2}\nabla^2\vv)$ with $p>d$
and $\nu=\nu(\theta)>0$, has been used in isothermal situations with inelastic
deformation without diffusion and without inertia in \cite{Roub22QHLS}
and with diffusion but without inelastic deformation in \cite{RouSte23VESS},
while an anisothermal model without diffusion and without inelastic deformation
in \cite{Roub24TVSE} and with inelastic deformation in \cite{RouTom24IFST}.
In addition, another dissipative gradient term can be used for $\Lp$,
which expands the kinetic equation \eq{system-diffusion-eq} by a Laplace-type term. 

The a-priori estimation strategy (to be applied for a suitable Galerkin
semi-discretization such as in \cite{Roub22QHLS,Roub24TVSE,RouSte23VESS,RouTom24IFST})
is to use first the total-energy balance \eq{tot-engr}, which gives a uniform-in-time
$L^1(\varOmega)$-estimate of $\int_\varOmega\frac12\varrho|\vv|^2+e\,\d\xx$.

Then, one should estimate the dissipative terms. Actually, if it were no
diffusion, one can use the dissipation-energy balance arising from the sole
mechanical part of the system, i.e.\ from (\ref{the-system}a-c,e) tested
successively by $\frac12|\vv|^2$, $\vv$ (while using also \eq{evol-of-E++}),
and $\Lp$, which would give an $L^1$-estimate for the dissipation rate
which would be then use in the $L^1$-theory for the heat equation to obtain
an estimate for $\nabla\theta$. Yet, with diffusion, it must to be more
tricky. To this aim, one should
use the entropy balance \eq{ent-balance} which, on the other hand, can yield
directly an estimate on $\nabla\theta$ under suitable assumption. The last
point, developed in the context of viscoelastic fluids in
\cite[Sect.\,2.2.3]{FeiNow09SLTV}, seems an advantageous technique even
if diffusion would not be involved. More specifically, we integrate
\eq{ent-balance} in time with using \eq{data-r} to obtain
\begin{align}\nonumber
\int_0^T\!\!\!\int_\varOmega\frac{
\bbD_\mathrm{visc}\ee(\vv){:}\ee(\vv)\!}\theta+\frac{[R_{\rm plast}]_{\Lp}'\!\big(\ZETA,\theta;{\Lp}/\theta\big){:}\Lp\!\!\!}\theta
+\nabla\frac\mu\theta{\cdot}\bbK_\mathrm{diff}\nabla\frac\mu\theta 
+\nabla\frac1\theta{\cdot} \bbK_\mathrm{heat}\nabla\frac1\theta
\,\d\xx\d t
\\[-.5em]\nonumber
=\int_\varOmega\psi_\theta'(\Fe(0),\ZETA(0),\theta(0))-\psi_\theta'(\Fe(T),\ZETA(T),\theta(T))\,\d\xx<+\infty\,.
\end{align}
Under appropriate  assumptions (in particular, assuming a polynomical-type
growth of the heat capacity rather than a constant heat
capacity), the right-hand side can be shown bounded
by using the already proved apriori estimates. Then, assuming
suitable growth of $\bbD_\mathrm{visc}=\bbD_\mathrm{visc}(\theta)$,
$\bbK_\mathrm{diff}=\bbK_\mathrm{diff}(\theta)$, and
$\bbK_\mathrm{heat}=\bbK_\mathrm{heat}(\theta)$, we can obtain a-priori estimates
of $\ee(\vv)$, $\nabla\mu$, and $\nabla\theta$. Analogously, also $\Lp$ can be
estimated under suitable assumption on $R_{\rm plast}$. The mentioned multipolar
modification can yields an estimate on $\nabla^2\vv$.

The last mentioned estimate would imply that the velocity field $\vv$ is
Lipschitz continuous in space, which in turn guarantees
a certain regularity of $\Fe$ and $\varrho$, including bounds on
$\nabla\Fe$ and positivity of $\det\Fe$ and $\varrho$. 
Assuming uniform convexity of $\psi(\Fe,\cdot,\theta)$,
from $\nabla\mu$ and $\nabla\Fe$ and also $\nabla\theta$ we can estimate
$\nabla\ZETA$; realize that $\mu=\psi_\ZETA'(\Fe,\ZETA,\theta)$
so that
$$
\nabla\ZETA=\frac{\nabla\mu-\psi_{\Fe\ZETA}''(\Fe,\ZETA,\theta)\nabla\Fe-
\psi_{\ZETA\theta}''(\Fe,\ZETA,\theta)\nabla\theta)}{\psi_{\ZETA\ZETA}''(\Fe,\ZETA,\theta)}\,.
$$
The details concerning the above estimated and relevant assumption
on the data are rather technical and we avoid specifying them.

All these kinds of estimates would allow for a limit passage in a
suitable (semi) discretization constructed similarly as in
\cite{Roub22QHLS,Roub24TVSE,RouSte23VESS,RouTom24IFST}. In such
a way, one could prove the existence of suitably defined weak solutions
of an initial-boundary-value problem on a finite time interval $[0,T]$.

\appendix 
\section{Appendix: Jacobi's identity for Poisson operators with block structure} 
\label{app:PoissBlock}

In many applications, the Poisson operator $\bbJ$ has a block structure on the
product space $X=X_1\ti \cdots \ti X_N$, namely for $q=(q_1,,q_2,\ldots, q_N )$ we have
\begin{equation}
  \label{eq:PoissBlock}
  \bbJ(q) = \bma{ccccc}
\bbJ_{11}(q_1) & \bbJ_{12}(q_2)& \cdots&  \bbJ_{1N}(q_N)\\
\bbJ_{21}(q_2) &  0 & \cdots&  0\\
\vdots & \vdots& \ddots &  \vdots \\
\bbJ_{N1}(q_N) & 0 & \cdots&  0
 \ema \quad \text{with }\ \bbJ_{n1}(q_n)^*=-\bbJ_{1n}(q_n)\,. 
\end{equation}
Note that the operators $\bbJ_{n1}$ and $\bbJ_{1n}$ are allowed only to depend on the
component $q_n$. The following result provides necessary and sufficient
conditions for such $\bbJ$ to satisfy Jacobi's identity 
\begin{equation}
  \label{eq:Jacobi.Ident}
 \big\langle \zeta_1,
\rmD\bbJ(q)[\bbJ(q)\zeta_2]\zeta_3\big\rangle+\text{cycl.perm} \equiv 0 \ 
\text{ for all } q\in X \text{ and all } \zeta_1,\zeta_2,\zeta_3 \in X^*.
\end{equation}

\begin{proposition}[Jacobi's identity]
\label{pr:Jacobi:Ident} 
Assume that $X$ has the block structure $X=X_1\ti \cdots \ti X_N$ for some
$N \geq 2$ and assume that $\bbJ$ satisfies \eqref{eq:PoissBlock}. Then, $\bbJ$
satisfies Jacobi's identity \eqref{eq:Jacobi.Ident} if and only if the
following conditions hold:
\begin{subequations}
\label{eq:Conds.JI.all}
\begin{align}
&\label{eq:Conds.JI.A} 
\bbJ_{11} \text{ satisfies Jacobi's identity on } X_1; \\
&\label{eq:Conds.JI.B}  \rmD\bbJ_{11}(q_1)= \rmD\bbJ_{11}(0) =:
\bbA:X_1\to \mathrm{Lin}_\text{\rm skw}(X_1^*,X_1), \text{ i.e. } \bbJ_{11}(q_1)
=\bbJ_{11}(0)+ \bbA q_1; \\
&\nonumber  \text{for } n \in \{2,\ldots,N\} \text{ and all }
\zeta\in X_n^*,\ v,w \in X_1^*,\ q_n\in X_n \text{ we have the
  identity}\\
& \label{eq:Conds.JI.C} 
\ \big\langle v, \big(\bbA \bbJ_{1n}(q_n) \zeta\big) w \big\rangle_{X_1} \!\!
= \big\langle \zeta, \rmD\bbJ_{n1}(q_n)\big[\bbJ_{n1}(q_n)v\big] w \big\rangle_{X_n} 
\!\!\!-\big\langle \zeta, \rmD\bbJ_{n1}(q_n)\big[\bbJ_{n1}(q_n)w\big] v
\big\rangle_{X_n}\!. 
\end{align}
\end{subequations}
\end{proposition}

\noindent{\it Proof.}
Writing $\zeta^\ALPHA=(\zeta^\ALPHA_1,\ldots, \zeta^\ALPHA_N) \in X_1^*\ti
\cdots \ti X^*_N$ and using the special form of $\bbJ$ in
\eqref{eq:PoissBlock} gives 
\begin{align*}
\big\langle \zeta^1, \rmD\bbJ(q)[\bbJ(q)\zeta^2]\zeta^3\big\rangle &= 
 \big\langle \zeta^1_1, \rmD\bbJ_{11}(q_1)[\bbJ_{11}(q_1)\zeta^2_1]\zeta^3_1\big\rangle
+ \sum_{n=2}^N \Big( 
 \big\langle \zeta^1_1, \rmD\bbJ_{11}(q_1)[\bbJ_{1n}(q_n)\zeta^2_n]\zeta^3_1\big\rangle
\\[-.5em]
&\quad+\big\langle\zeta^1_1,\rmD\bbJ_{1n}(q_n)[\bbJ_{n1}(q_n)\zeta^2_1]\zeta^3_n\big\rangle
+\big\langle\zeta^1_n,\rmD\bbJ_{n1}(q_n)[\bbJ_{n1}(q_n)\zeta^2_1]\zeta^3_1\big\rangle\Big).
\end{align*} 

Adding cyclic permutations and considering first $\zeta^\ALPHA_n=0$ for
$n\geq2$, we see that it is necessary that $\bbJ_{11}$ satisfies Jacobi's
identity on $X_1$, which is the assumption \eqref{eq:Conds.JI.A}.   

The remaining terms are linear in $\big(\zeta^\ALPHA_n\big)_{\ALPHA=1,2,3,\,n\geq 2}$
and bilinear in $\zeta^\ALPHA_1$. Fixing one $n \in \{2,\ldots,N\}$ and setting
$\zeta^\ALPHA_m=0$ for $m \not\in\{1,n\}$,  we obtain a linear expression in
$\zeta^\ALPHA_n$ for each $\ALPHA \in \{1,2,3\}$. Considering all cyclic
permutations this leads to the three terms that have to sum up to $0$. Choosing
one $\ALPHA$ and denoting by $\BETA$ and $\GAMMA$ the other two indices, the
condition reads
\begin{align*} 
0= \big\langle \zeta^\BETA_1, \rmD\bbJ_{11}(q_1)[\bbJ_{1n}(q_n)\zeta^\ALPHA_n]\zeta^\GAMMA_1
 \big\rangle + \big\langle \zeta^\GAMMA_1,
\rmD\bbJ_{1n}(q_n)[\bbJ_{n1}(q_n)\zeta^\BETA_1]\zeta^\ALPHA_n \big\rangle 
+ \big\langle \zeta^\ALPHA_n,
\rmD\bbJ_{n1}(q_n)[\bbJ_{n1}(q_n)\zeta^\GAMMA_1]\zeta^\BETA_1 \big\rangle\,.
\end{align*}
Since the first term depends on $q_1$ but not the other two, this relation can
only hold if $\rmD \bbJ_{11}$ is independent of $q_1$, which provides
the condition \eqref{eq:Conds.JI.B}. 

Moreover, using $\bbJ_{1n}=- \bbJ_{n1}^*$ we see that the last condition is the
same as \eqref{eq:Conds.JI.C}. Hence, the necessity of \eqref{eq:Conds.JI.all}
is established. However, by construction the sufficiency is clear.
\mbox{}\hfill$\Box$

\bigskip\bigskip

\baselineskip=13pt
{\small
\noindent{\it Acknowledgments:}
A.M.\ acknowledges partial support by the Deutsche Forschungsgemeinschaft
through Collaborative Research Center SFB 1114 ``Scaling Cascades in Complex
Systems'' (Project Number 235221301).  T.R.\ acknowledges hospitality and a
support from the Weierstrass institut Berlin. This research was also supported
from the CSF project no.\,23-06220S, and from the institutional support
RVO:\,61388998 (\v CR).}


\def\cprime{$'$}

\end{sloppypar}

\end{document}